\documentclass{article}

\usepackage{arxiv}
\makeatletter
\renewcommand{\normalsize}{%
  \fontsize{12pt}{14pt}\selectfont  %
  \abovedisplayskip      7\p@ \@plus 2\p@ \@minus 5\p@
  \abovedisplayshortskip \z@ \@plus 3\p@
  \belowdisplayskip      \abovedisplayskip
  \belowdisplayshortskip 4\p@ \@plus 3\p@ \@minus 3\p@
}
\makeatother

\RequirePackage{amsmath,amsfonts,amssymb}
\RequirePackage[authoryear]{natbib}

\usepackage{snapshot}
\usepackage{graphicx}
\usepackage{enumerate}
\usepackage{comment}
\usepackage{url} 
\RequirePackage[colorlinks,citecolor=blue,urlcolor=blue]{hyperref}

\usepackage{bm, amssymb}
\usepackage{wrapfig}
\usepackage{dsfont}

\usepackage[ruled,vlined]{algorithm2e}

\usepackage{phyloHPYP}  
\usepackage[noabbrev]{cleveref}
\usepackage[normalem]{ulem}

\newtheorem{example}{Example}

\title{Detecting Jumps on a Tree: a Hierarchical Normalized Stable Process Model for Evolution of Discrete Distributions}

\renewcommand{\shorttitle}{Hierarchical \bnpfull\  on trees}

\usepackage{authblk}

\author[1]{Hanxi Sun}
\author[2]{Heejung Shim$^\ast$}
\author[1]{Vinayak Rao$^\ast$}
\affil[1]{Department of Statistics, Purdue University \par
\vspace{-.06in} \texttt{hanxi.sun.work@gmail.com, varao@purdue.edu}
\vspace{.1in}}
\affil[2]{School of Mathematics and Statistics and Melbourne Integrative Genomics, University of Melbourne \par
\vspace{-.06in}  \texttt{heejung.shim@unimelb.edu.au}}

\date{}

\begin{document}

\maketitle
\vspace{-.6in}
\begin{center}
\small $^\ast$Equal contribution, co-supervised this work.
\end{center}
\vspace{.2in}


\begin{abstract}
    This work focuses on clustering populations whose hierarchical dependency structure can be described by a tree. 
    A particular example that is the focus of this work is a phylogenetic tree, whose nodes represent different biological species. 
    In this problem, clustering the populations at the leaves of the tree is equivalent to identifying branches in the tree where the populations at the parent and child node have significantly different distributions. 
    We construct a nonparametric Bayesian model based on the hierarchical \bnpfull\  and the Poisson process to exploit this hierarchical structure, with a key contribution being the ability to share statistical information between subpopulations. We develop an efficient particle MCMC algorithm to address computational challenges involved with posterior inference.
    We illustrate the efficacy of our proposed approach on both synthetic and real-world problems.
\end{abstract}



\section{Introduction} \label{sec:introduction}
Modern datasets often exhibit rich underlying structure arising from the mechanistic, spatial, or temporal processes that generated them. Trees are an example of such structure, representing observations in a hierarchy of partially overlapping sets at multiple granularities. A classic example, and the focus of this work, is a phylogenetic tree describing the evolutionary relationships among species. Although our motivating examples are phylogenetic, we take a broader view and also consider evolving languages and other social norms on trees. Accounting for the underlying tree structure allows statistical strength to be shared across related observations while permitting variation across different parts of the tree.

In phylogenetic comparative analyses, many approaches
assume a single phenotype for each species and model its evolution as either gradual, through a diffusion process such as Brownian motion~\citep{felsenstein1985, freckleton2002phylogenetic, brawand2011evolution}, or abrupt, through a sequence of jumps modeled by a compound Poisson process or the pure-jump component of a Lévy process~\citep{landis2012phylogenetic, landis2017punctuated, duchen2017inference}. 
In practice, multiple phenotype observations may be available for each species, exhibiting variation due to measurement error as well as genetic and environmental effects~\citep{felsenstein2008, revell2012}. Several approaches   explicitly account for intraspecific variation by associating each species with a probability distribution whose \emph{parameters} evolve along the tree, either continuously or through jumps~\citep{ives2007, felsenstein2008, revell2012, kostikova2016bridging, ansari2016bayesian}. 
In this paper, we model such jumps,  allowing us both to test phylogenetic trait association, and to estimate a segmentation of the tree.

Parametric models offer an interpretable and computationally convenient approach, but impose strong assumptions on the node-specific distributions. Nonparametric Bayesian models~\citep{ghosal2017fundamentals} provide greater flexibility by placing priors directly on infinite-dimensional probability distributions, allowing complex, multimodal and otherwise non-standard phenotype distributions. The approach of~\citet{ansari2016bayesian} takes a step in this direction for categorical data, though each jump generates a new finite-dimensional distribution from a Dirichlet distribution. 

In this work, we develop a fully nonparametric Bayesian model for evolving distributions on a phylogenetic tree.  Our nonparametric building block is the \bnpfull\ (\bnp)~\citep{kingman1975random, regazzini2003distributional}, closely related to the Pitman--Yor process~\citep{pitman-yor}. A convenient marginalization property allows us to integrate out intermediate clusters, yielding a computationally tractable model. 
Ours is a hierarchical model that centers the new distribution at the old one, encoding the prior expectation that neighboring distributions should be similar. This enables statistical sharing across the tree, and contrasts with~\citet{ansari2016bayesian}, where the distribution after a jump is drawn independently of the preceding distribution. 
We discuss the implications of this in~\Cref{sec:treebreak}.
We apply our method to real datasets with binary and nonbinary phenotypes, showing improved performance relative to~\citet{ansari2016bayesian}. Our method is implemented in Python3 and is available at \url{https://github.com/varao/jumpsontree}.

\subsection{Related work}
Our approach is related to several lines of work on Bayesian nonparametrics, hierarchical clustering, changepoint detection, and phylogenetic trait association. From the Bayesian nonparametric perspective, our model builds on hierarchical and dependent priors~\citep{teh2010hierarchical}, with our use of the hierarchical normalized stable process~\citep{regazzini2003distributional, camerlenghi2019distribution} inspired by marginalization constructions developed for language modeling~\citep{teh2006hierarchical}. Unlike these settings, our hierarchy is given by a phylogenetic tree, with the locations of distributional changes inferred from the data, rather than given. This also distinguishes our problem from classical hierarchical clustering and Bayesian tree-valued models~\citep{heller2005bayesian, neal2003density}, where the hierarchy itself is inferred and observations are exchangeable \emph{a priori}. More broadly, our model can be viewed as a nonparametric Bayesian changepoint model~\citep{smith1975bayesian, muliere1985change, mira1996bayesian}, with change points occurring on a fixed tree rather than on the real line.

Several existing methods specifically address phylogenetic trait association or change detection. Most closely related to our work is {\tt treeBreaker}~\citep{ansari2016bayesian}, which we discuss in~\Cref{sec:treebreak} after specifying our model. \texttt{treeSeg}~\citep{behr2020testing} formulates phylogenetic trait association as a multiscale change-point problem, using a constrained optimisation based on the multiscale SMUCE statistic. 
It returns an estimated segmentation and associated confidence sets and comes with  a tree-wide false-positive guarantee: the probability of including any spurious active edge is at most the user-specified significance level $\alpha$. We discuss this more in our empirical evaluations, and in the supplementary material.

Like our work, {\tt CRP-Tree}~\citep{zhang2024crp} models phylogenetic trait association using a Chinese Restaurant Process--inspired model, in which a parameter controls the degree of preferential attachment. Under the null, there is no preferential attachment, corresponding to a coalescent null model. Significance is assessed using the number of same-type attachments and a permutation test. While {\tt CRP-Tree} provides a global test of phylogenetic association, it does not return an estimated segmentation. This is also true of parsimony-based approaches such as the association index of~\citet{slatkin1989cladistic}, and {\tt BaTS}~\citep{parker2008correlating}, which extends such approaches to account for phylogenetic uncertainty.

\section{Problem setup and modeling approach} \label{sec:model}


Let $T=(V,E)$ be a known rooted tree, with $V = \{0, 1, \dots, |V|-1\}$ the node set, and $E$ the edge set. 
We reserve index $0$ for the root; the remaining indices are purely labels and do not imply any ordering of the nodes. Each node $v\in V$ indexes a population with $\N_v\geq0$ observations on a common finite space $\cX=\{1,\dots,K\}$.
Write $\obs{v}{i}$ for the $i$th  measurement from population $v$, with our observed dataset being $\allObs=\cbr{\obs{v}{i}:v\in V,\ i=1,\dots,\N_v}$. 
Each edge $e\in E$ has length $\ell(e)>0$, representing the expected dissimilarity between the distributions of observations at its parent and child nodes. Treating each edge as a line segment of length $\ell(e)$ gives a continuous tree. The distance between any two locations $s$ and $t$ on this tree -- each either a leaf, an internal node, or a point partway along a branch -- is
$\operatorname{dist}(s,t)= \sum_{e \in E} \ell(e\cap \pth(s,t))$, where $\pth(s,t)$ is the unique path connecting $s$ and $t$.

\begin{example}[binary trait -- HLA allele data,~\Cref{sec:exp-hla}] \label{eg:virus}
Here $\mathcal{X} = \{0,1\}$, where $x_{(v,i)}=1$ indicates that the $i$-th sampled host at node $v$ of the virus phylogenetic tree carries the HLA allele B57, and $x_{(v,i)}=0$
otherwise. The tree $T$ is a phylogenetic tree inferred from a 1000-nucleotide segment of the aligned viral genomes, with its topology and branch lengths reflecting the evolutionary relationships among the sampled viral strains.  Node $v$ corresponds to a clade of hosts, and $N_v$ is the number of hosts sampled at that clade; observations are available only at the 261 leaves.
\end{example}

\begin{example}[four-category trait -- Uto-Aztecan post-marital residence data,~\Cref{sec:exp-pmr}]
\label{eg:fourcateg}
Here $\mathcal{X} = \{\text{patrilocal}, \text{matrilocal}, \text{ambilocal}, \text{neolocal}\}$, $x_{(v,i)}$ records the predominant post-marital residence practice reported for the $i$-th observation at language community $v$ in the Uto-Aztecan language tree. The tree $T$ here is a language tree, with its topology reflecting the historical relationships among Uto-Aztecan language communities rather than biological descent. Observations are available at the 26 leaves of the tree, one per language community.
\end{example}

In both examples, $N_v = 0$ for internal nodes and $N_v > 0$ only at the leaves. 
Our model also accommodates $N_v > 0$ at internal nodes, e.g.\ when fossil measurements are available for a common ancestral node of two or more leaf nodes.
We model observations at node $v$ as i.i.d. draws from a node-specific probability measure $G_v$ on $\mathcal{X}$:
$$
x_{(v,i)} \mid G_v \overset{\text{i.i.d.}}{\sim} G_v, \qquad i = 1, \dots, N_v.
$$
The $G_v$'s themselves are random, with a hierarchical prior across nodes described later. Importantly, the distributions $G_v$ and $G_u$ at two distinct nodes may or may not coincide, and our goal is to detect the branches along which they differ (corresponding to \emph{jumps}).

\subsection{Poisson process jumps on the tree}
\label{sec:poiss_tree}
We model changes in the distribution across the tree as occurring at a random, finite set of \emph{jump points} $J$, realized from a Poisson process on the continuous tree defined above. Throughout, ``depth" refers to $\dist(0,\cdot)$, the cumulative branch length from the root. 

The simplest choice is a homogeneous Poisson process with rate $\lambda > 0$, which assigns the same jump intensity to every unit of branch length. 
Because the number of branches typically grows approximately exponentially with depth, the aggregate jump intensity also grows with depth, favoring jumps near the leaves purely as an artifact of tree shape. 

Instead, we calibrate the Poisson rate so that the aggregate jump intensity is constant across depth. Concretely, if $k(p)$ branches are present at depth $p$, we set the Poisson rate  for each branch at that depth to $\lambda/k(p)$, so the aggregate rate remains $\lambda$. Equivalently, we partition branches into segments over which the number of branches is constant, and rescale each segment by the reciprocal of this number, and place a homogeneous rate-$\lambda$ Poisson process on the resulting tree. Details are given in the supplementary material. The rate $\lambda$ can be fixed a priori or given an $\mathrm{Exponential}(\rho)$ prior and learned from the data. 

We also consider a simpler approximation that rescales each entire branch by a single factor based on the number of leaves in its subtree, rather than partitioning branches into segments. Our model's performance is robust to both choices.

\subsection{Node distributions and pruning}
\label{sec:pruning}
With the Poisson jump set $J$ fixed, let $\bar J := J \cup \{0\}$: we treat the root as a distinguished jump point at which the process is initialized.
In~\Cref{sec:hnsp} we define an HNSP construction that assigns a random distribution $G^*_j$ to every $j \in \bar J$. 

For any location $s$ on the tree, let $\nearjmp(s)$ denote the last point of $\bar J$ encountered along the path from the root to $s$. 
Define $G_s := G^*_{\nearjmp(s)}$. In particular, the distribution associated with node $v \in V$ is $G_v := G^*_{\nearjmp(v)}$.
Thus, $\pth(u,v) \cap J  = \varnothing$ implies $G_u = G_{v}$.
Our Poisson construction of $J$ allows multiple jumps on the same branch, in which case the child node is assigned the distribution at the last jump. Under our HNSP model, this captures progressively larger distributional changes along that branch.

Define an equivalence relation $\sim_J$ on $V$ by $u \sim_J v \Longleftrightarrow \pth(u,v) \cap J = \varnothing$. 
Each equivalence class is a connected, jump-free subtree whose nodes share the same distribution. We collapse each class into a single node, writing $V_J$ for the resulting node set. Observations from all nodes in a class are then associated with the corresponding node in $V_J$ and modeled as i.i.d. draws from its shared distribution.

\begin{figure}
	\includegraphics[width=\linewidth]{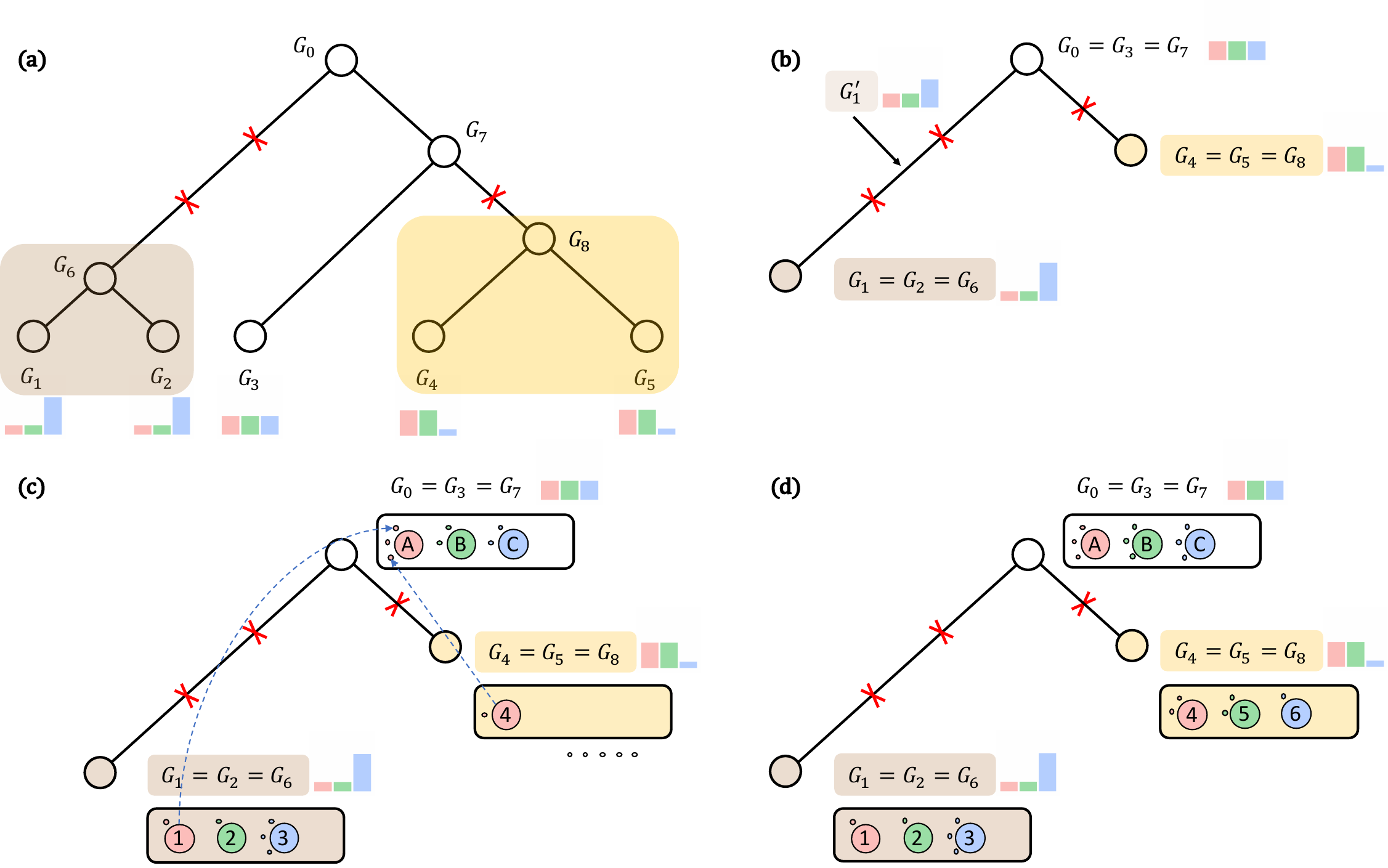}
    \caption{An example of the partition of tree nodes induced by jumps (red crosses) and the corresponding probability distributions.
\textbf{(a)} Color-coded groups of nodes. Bar plots show the distributions associated with the corresponding leaf nodes; leaves in the same group share the same distribution.
\textbf{(b)} The pruned tree induced by the jumps in (a), with each group represented by a single node.
\textbf{(c)} The color-coded clustering of the first 6 observations, along with the coupling of restaurants across nodes under the CRF. Tables 1 and 4, in different restaurants, are seated at a single table A in the root, and so share the same value (see \Cref{sec:hpyp,sec:crf} and the supplementary material).
\textbf{(d)} The clustering structure of all 10 observations.}
	\label{fig:tree_and_crf}
\end{figure}

Each equivalence class in $V_J$ has a unique entry point in $T$ (except the class containing the root), with edges crossing between classes in one-to-one correspondence with the jump-carrying edges of $T$. We connect two nodes of $V_J$ whenever an edge of $T$ connects their corresponding classes, and write $E_J$ for the resulting edge set. We call the resulting tree  $T_J=(V_J,E_J)$ the \emph{pruned tree}; see~\Cref{fig:tree_and_crf}(a) and (b), and the supplementary material.

Branch lengths affect the construction only through $J$, and 
$T_J$ enters the remainder of the construction through its topology and per-edge jump counts alone. 
Because multiple jumps may occur on the same branch, the cardinality of $V_J$ can be smaller than that of $J$.

\subsection{The hierarchical normalized stable process (HNSP) } \label{sec:hnsp}

Throughout, $G$ (without a subscript) denotes a generic random probability measure on $\cX$. 
The Pitman--Yor process~\citep{pitman-yor}, parametrized by a concentration parameter $\alpha$, a discount parameter $\d\in(0, 1)$ and a base distribution $\H$ on $\cX$, is a nonparametric prior over discrete probability distributions $G$ on $\cX$:
$$\given{\G}{\alpha, \d, \H}\sim\PYP{\alpha, \d, \H}.$$
Setting $d=0, \alpha > 0$ recovers the well-known Dirichlet process (DP)~\citep{ferguson1973bayesian}, while $\alpha=0, d \in (0,1)$ yields the \bnpfull\ $\NSP{d,H}$~\citep{lijoi2014bayesian}. 
The base distribution $H$ is the prior mean, with $\EE[\G] = H$, while the discount parameter $d$ governs the variability around this mean. 
Under the usual support condition $G_{\textrm{data}} \ll H$, the NSP assigns positive mass to every Kullback-Liebler neighborhood of any data-generating distribution $G_{\textrm{data}}$.  We focus on the NSP because it is simpler than the PYP, and because it possesses a convenient marginalization property that substantially simplifies posterior inference. 

We now specify the probability distribution $G^*_j$ associated with each jump $j \in \bar J$. 
Starting with the root, we take $H$ to be  uniform and set
$G^*_0 \mid d \sim \text{NSP}(d, H)$.   For each jump point $j \in J$, let $\parent(j) \in \bar J$ denote its immediate predecessor jump point or parent. We then recursively generate the associated distribution as
$G^*_j \mid d, G^*_{\parent(j)} \sim \text{NSP}(d, G^*_{\parent(j)})$.
Thus, each child distribution is centered at its parent's distribution rather than at a fixed $H$, allowing distributional changes to accumulate along successive jumps. This recursively defines the hierarchical normalized stable process (HNSP) on $T_J$. We write this as
$$G^*_J := \{G^*_j, \,\, j \in \bar J\} \sim \operatorname{HNSP}(d,H).$$

Combined with the Poisson prior on 
$J$, the HNSP prior induces a probability measure at every node $v \in V$ through the assignment $G_v:=G_{\nearjmp(v)}^*$, giving the collection $G_T:=\{G_v:v\in V\}$.
Conditional on $G_T$, the observations at any node $v$ are i.i.d.\ samples from its associated probability distribution $G_v$. Given observations $\allObs$, inference amounts to characterizing the posterior over $J$ and $G^*_J$ (which together induce $G_T$). 



\subsection{Relation to \texttt{treeBreaker}~\citep{ansari2016bayesian}} \label{sec:treebreak}
\texttt{treeBreaker} also models jumps on the tree as a Poisson process. However, in this model, each jump is assigned an independent Dirichlet-distributed distribution with fixed hyperparameters, rather than a distribution centered on its parent distribution.
Consequently, because the $G^*_j, j \in \bar J$ are independent \emph{a priori}, posterior inference for any $G^*_j$ is informed only by observations from nodes belonging to the corresponding equivalence class. When few such observations are available, this can result in less stable posterior estimates. 

Our hierarchical model borrows statistical strength across all nodes in the tree, with dependence between distributions decaying with tree distance and, conditional on $J$, with the number of intervening jumps. Thus, observations from nearby nodes can inform inference even when data in an individual node are sparse.
As we show empirically, borrowing strength also allows our method to detect small, aligned distributional changes more reliably.


\subsection{Marginalization properties of HNSP} \label{sec:hpyp}
We discuss two useful marginalization properties of the HNSP that let us integrate out 1) the distributions at intermediate jumps, and 2) the infinite-dimensional distributions associated with the remaining vertices of the pruned tree via the Chinese restaurant franchise (CRF).  

\subsubsection{Marginalizing out intermediate jumps}
 Consider an edge with two jumps, $j$ and its parent $j'=\parent(j)$. 
We call $j'$ an \emph{intermediate} jump: this satisfies $j' \neq \nearjmp(v)$ for every 
$v \in V$. 
Let $j'' := \parent(j')$. 
For the NSP, the following marginalization holds~\citep{wood2009stochastic}:
$$G^*_j \mid d, G^*_{j''} \sim \NSP{d^2, G^*_{j''}}.$$
Thus, the intermediate distribution $G^*_{j'}$ can be integrated out. 
For $b$ consecutive jumps $j_1, \dots, j_b$ on a branch, ordered from root to leaf, $j_1, \dots, j_{b-1}$ are all intermediate jumps, and
$$G^*_{j_b} \mid d, G^*_{\parent(j_1)} \sim \NSP{d^b, G^*_{\parent(j_1)}}.$$
Thus, by keeping track of the number of jumps along each branch, we can marginalize out intermediate distributions. 
Distributions at jumps that immediately precede nodes in $V$
 are instantiated when either (i) the corresponding vertex of $T_J$ has multiple children, since it then serves as a shared ancestor inducing dependence among its descendants, or (ii) it carries observations.
Consequently,  we instantiate one distribution per vertex of the pruned tree $T_J$, whether or not it carries observations, without sampling distributions at intermediate jump points.
This reduces the computational cost when multiple jumps occur on the same branch. 

\citet{ansari2016bayesian}'s {\tt treeBreaker} does not encounter this issue because their model draws each new distribution independently of its predecessor, making multiple jumps along a branch equivalent to a single jump. 
We discussed the statistical cost of this in~\Cref{sec:treebreak}.

\subsubsection{The Chinese restaurant franchise}
\label{sec:crf}

The second marginalization property concerns the remaining measures $G^*_{T_J} := \{G^*_v : v \in V_J\}$ on the pruned tree $T_J$. For an NSP, each of these is infinite-dimensional, and thus cannot be represented exactly on a computer. Instead, we marginalize these measures using a Chinese restaurant franchise (CRF)~\citep{teh2010hierarchical,teh2006hierarchical,teh2005HDP}, yielding a finite representation for the observed data.

We first recall the Chinese restaurant process for a single NSP.
If $G^* \sim \NSP{d,\H}$, then i.i.d.\ observations $(x_1,\dots,x_n)$ from $G^*$ can be generated without sampling $G^*$ as follows: 
\begin{itemize}
    \item Simulate the first observation from the base measure: $x_1 \sim H$
\item Sequentially assign each new observation to an existing cluster or create a new one:
\begin{equation}
	\label{eqn:crp_maintext}
	\given{x_{i+1}}{\{x_1, \dots, x_i\},~d,~H} 
	\sim \sum_{k=1}^{K_i} \frac{\n_k-\d}{ i}\delta_{x^*_k} + \frac{K_i\d}{i} H.
\end{equation}
Here $K_i$ denotes the number of clusters formed by the first $i$ observations, and $\n_k$ and $x^*_k$ are the cluster size and the first observation of cluster $k$, respectively.
\end{itemize}

For the HNSP, the CRF associates a restaurant with each node $v_J$ of $T_J$, populated by data from all nodes in the equivalence class of $v_J$. 
Henceforth, nodes and parent relationships refer to $T_J$, and cluster counts and sizes in a restaurant are aggregated over its class.

To generate a new observation $x_{(v,i)}$ at node $v$, we first assign it to a cluster $\tbl{v}{i}$ in its restaurant
according to \Cref{eqn:crp_maintext}.
If it joins an existing cluster, its value is inherited from that cluster. If it starts a new cluster, its value must instead be drawn from the parent distribution $G^*_{\parent(v)}$. We treat the new cluster as a customer in the parent restaurant and recursively apply the same rule. The upward recursion terminates when an existing cluster is reached, whose value is inherited, or when a new cluster is created at the root, whose value is drawn from $H$. Thus, each observation is represented by a finite cluster-assignment sequence $C_{(v,i)}$ along the path to the root, without explicitly sampling any of the $G^*_v$. Note that while the HNSP proceeds from the root downward as a prior over distributions, the CRF, which integrates these distributions out, proceeds from observation nodes upward to the root; see~\Cref{alg:crf}.
\begin{algorithm}[t]
\caption{CRF generation of observation $i$ at node $v$ (node $v_J$ in $T_J$)}
\label{alg:crf}
\SetKwInOut{Input}{Input}\SetKwInOut{Output}{Output}
\Input{Current seating of all restaurants on the path in $T_J$ from $v_J$ to the root}
\Output{Value $x_{(v,i)}$ and cluster-assignment sequence $C_{(v,i)}$}
$u \leftarrow v_J$;\quad $C_{(v,i)} \leftarrow ()$  \tcp*{counts and tree relations are on $T_J$, see text}
\Repeat{$c$ is an existing cluster \textbf{or} $u$ is the root}{
  Draw a cluster $c$ in restaurant $u$ from \eqref{eqn:crp_maintext}; append $c$ to $C_{(v,i)}$\;
  \lIf{$c$ is a new cluster and $u \neq \mathrm{root}$}{$u \leftarrow \mathrm{pa}(u)$}
}
$x_{(v,i)} \leftarrow$ value of $c$ if $c$ is an existing cluster, else a fresh draw from $H$\;
\Return $(x_{(v,i)}, C_{(v,i)})$\;
\end{algorithm}

The complete generative process for the observations $\allObs=\cbr{\obs vi}$ is as follows:
\begin{enumerate}
	\item Rescale the branch lengths of the tree $T$ so that jumps occur at a constant rate over depth.
	\item Fix or sample the Poisson intensity $\lambda$. We use the weakly informative $\mathrm{Exponential}(\rho)$ prior.
\item Sample the jump configuration $J$ from a rate-$\lambda$ Poisson process on the rescaled tree; in practice (Section~\ref{sec:hpyp}), record only the branch counts, $\Jump = (\jump_1,\dots,\jump_{|E|})$.  
	\item Prune $T$ to obtain $T_J$. Each edge of $T_J$ corresponds to an edge $e$ of $T$ satisfying $\jump_e \geq 1$, while its vertices correspond to groups of populations in $T$ sharing the same distribution. 
    \item Specify the HNSP on $T_J$: for the root, $G^*_0 \sim \NSP{d,H}$, and for each non-root vertex $v_J$ with incoming edge $e_J$, $G^*_{v_J} \mid G^*_{\parent(v_J)} \sim \NSP{d^{\jump_{e_J}},G^*_{\parent(v_J)}}$. 
	\item Generate the observations sequentially using the CRF representation of the HNSP.
\end{enumerate}

\section{Posterior Inference} \label{sec:inference}
Given the observations $\allObs = \{x_{(v,i)} : v \in V,\ i = 1,\dots,N_v\}$, our goal is to compute the posterior distribution over the jump configuration and Poisson rate. We work with the branch-count representation $\Jump = (b_e)_{e \in E} \in \mathbb{Z}_{\geq 0}^{|E|}$ of the jump process introduced earlier, where $b_e$ is the number of jumps on edge $e$. This uniquely determines the pruned tree $T_J$ and the HNSP discount exponents $d^{b_e}$. If needed, the jump locations can be sampled uniformly along their associated branches.
The target posterior is
$$p(\Jump, \lambda \mid \allObs) \ \propto\ p(\allObs \mid \Jump)\, p(\Jump \mid \lambda)\, p(\lambda),$$
where $p(\Jump \mid \lambda) = \prod_{e \in E} \mathrm{Poisson}(b_e; \lambda \ell'(e))$ is a product of independent Poisson probabilities under the rescaled branch lengths $\ell'(e)$ of~\Cref{sec:poiss_tree}, and $p(\lambda) = \mathrm{Exponential}(\rho)
$. We sample from this posterior by a Markov chain Monte Carlo (MCMC) algorithm (\Cref{alg:particle_mcmc}) that alternates between updating $\Jump$ given $\lambda$ (\Cref{sec:mcmc_jumps}) and updating $\lambda$ given $\Jump$ (\Cref{sec:jump_rate}). The main difficulty we must overcome is that the marginal data likelihood $p(\allObs \mid \Jump)$,  obtained by summing over all latent cluster configurations of the CRF, is not available in closed form. 

\subsection{Updating the jump configuration}
\label{sec:mcmc_jumps}
At each MCMC iteration, given the current jump vector $\Jump$, we propose $\Jump^*$ from a proposal distribution $q(\Jump^* \mid \Jump)$, and accept it with probability $\min\{1, A(\Jump, \Jump^*)\}$, where
\begin{align}
    A(\Jump, \Jump^*) \ =\ \frac{p(\allObs \mid \Jump^*)}{p(\allObs \mid \Jump)}\cdot\frac{p(\Jump^* \mid \lambda)\, q(\Jump \mid \Jump^*)}{p(\Jump \mid \lambda)\, q(\Jump^* \mid \Jump)}. \label{eq:mh_acc}
\end{align}
We use two proposal types, chosen uniformly at random at each iteration:
\begin{description}
    \item[\itshape Resample proposal.] Select an edge $e \in E$ uniformly, and resample $b_e^* \sim \mathrm{Poisson}(\lambda \ell'(e))$ from its prior, leaving $b_{e'}^* = b_{e'}$ for all $e' \neq e$.
    \item[\itshape Swap proposal.] Select a non-root edge $e \in E$ uniformly, and let $e' = \mathrm{pa}(e)$ be its parent edge. Then set $b_e^* = b_{e'}, b_{e'}^* = b_e$, leaving all other entries unchanged.
\end{description}
 While the resample proposal alone yields an ergodic MCMC chain, moving a jump between adjacent branches may require passing through low-probability configurations containing either jumps on both branches (before the original jump is removed) or on neither (before the new one is added). 
 The swap proposal moves directly between two such configurations in a single step, substantially improving mixing.

The proposal and prior ratios in~\eqref{eq:mh_acc} are easy to evaluate. The only intractable term is the likelihood ratio $p(\allObs \mid \Jump^*)/p(\allObs \mid \Jump)$, which we estimate using the particle filter described next.

 \subsection{Estimating the marginal likelihood via particle filtering}
 \label{sec:pf}
Computing $p(\allObs \mid \Jump)$ requires marginalizing over all latent CRF cluster configurations, and is intractable. We construct a positive, unbiased estimator $\widehat p(\allObs \mid \Jump)$ and use it within the pseudo-marginal approach of~\citet{andrieu2009pseudo}. The resulting Markov chain evolves on the  augmented state space $(\Jump, \widehat p(\allObs \mid \Jump))$. At each iteration, a proposed jump configuration $\Jump^*$ is paired with a freshly drawn estimate $\widehat p(\allObs \mid \Jump^*)$, and the Metropolis--Hastings acceptance ratio is obtained from \Cref{eq:mh_acc} by replacing the exact likelihood ratio with
${\widehat p(\allObs \mid \Jump^*)}/{\widehat p(\allObs \mid \Jump)}.$
The stationary distribution over $\Jump$ remains exactly $p(\Jump \mid \allObs,\lambda)$~\citep{andrieu2009pseudo}.

We construct $\widehat p(\allObs \mid \Jump)$ using a sequential Monte Carlo (particle filtering) algorithm that exploits the sequential CRF representation. 
We approximate $p(\allObs \mid \Jump)$ using $S$ particles, each particle $s = 1,\dots,S$ carrying a complete cluster-assignment history $\mathbf{C}^s = \{C^s_{(v,i)}\}$, where $C_{(v,i)}$ is the sequence of cluster assignments associated with $x_{(v,i)}$ in the CRF (\Cref{sec:crf}).

We implement an auxiliary particle filter with the fully adapted proposal of~\citet{pitt1999filtering},  described in \Cref{alg:pf}. As each observation is processed, particles are propagated according to the CRF conditional, assigned an incremental importance weight, and resampled according to these weights. The product of the resulting predictive likelihood estimates gives the unbiased estimator $\widehat p(\allObs \mid \Jump)$. The unbiasedness of this estimator follows from the general sequential Monte Carlo theory of~\citet{andrieu2010particle}, applied to the sequential generative structure of the CRF.

\begin{algorithm}[t]
\caption{$S$-Particle filter estimate of $p(\allObs \mid \Jump)$}
\label{alg:pf}
\KwIn{Jump configuration $\Jump$;  observations $\allObs$ in a fixed order}
\KwOut{Unbiased estimate $\widehat p(\allObs \mid \Jump)$}
$T_J \leftarrow \textsc{Prune}(T,b)$ \tcp*{depends only on $\Jump$; shared by all particles}
Fix an ordering of the observations $x_{(v,1)}, x_{(v,2)}, \dots$,  and let $(v,i)^-$ and $(v,i)^+$ denote the observations immediately preceding and following $x_{(v,i)}$ in this order.

For all particles, initialize the CRF as $C^s_{(v_1,1)} = (1,\dots,1)$; set $\widehat p \leftarrow H(x_{(v_1,1)})$\;
\ForEach{remaining observation $x_{(v,i)}$, in order}{
  \For{$s \leftarrow 1$ \KwTo $S$}{
    Sample $C^s_{(v,i)}$ by conditioning the generative process in \Cref{alg:crf} on $x_{(v,i)}$:
    \[ p\!\left(C^s_{(v,i)}\,\middle|\, x_{(v,i)},\, C^s_{(v,i)^-},\, \Jump \right)
      \propto
      \underbrace{ p\!\left( x_{(v,i)} \,\middle|\, C^s_{(v,i)},\, C^s_{(v,i)^-} \right) }_\text{likelihood}
      \underbrace{ p\!\left( C^s_{(v,i)} \,\middle|\, C^s_{(v,i)^-},\, \Jump \right) }_{\text{CRF predictive prior}}
    \]
If $x_{(v,i)}$ is the final observation, set $w^s_{(v,i)^+}=1$; otherwise, set $w^s_{(v,i)^+}$ to the one-step-ahead predictive probability:
    \[
      w^s_{(v,i)^+} = \sum_{c_{(v,i)^+}}
        p\big(x_{(v,i)^+} \mid c_{(v,i)^+},\, C^s_{(v,i)},\, \allObs_{(v,i)}\big)\,
        p\big(c_{(v,i)^+} \mid C^s_{(v,i)},\, \allObs_{(v,i)},\, \Jump\big)
    \]
  }
  $\widehat p \leftarrow \widehat p \cdot \dfrac{1}{S}\sum_{s=1}^{S} w^s_{(v,i)^+}$ \tcp*{incremental likelihood factor}
  Resample $\{C^s_{(v,i)}\}_{s=1}^S$ with replacement, with probability proportional to $\{w^s_{(v,i)^+}\}$\;
}
\Return $\widehat p$\;
\end{algorithm}

\begin{algorithm}
	\caption{Particle Markov Chain Monte Carlo} \label{alg:particle_mcmc}
	
	\DontPrintSemicolon
	\SetKwProg{Fn}{Function}{:}{}
	\SetKwFunction{trim}{TrimTree}
	\SetKwFunction{pf}{ParticleFilter}
	\SetKwFunction{ParticleMCMC}{ParticleMCMC}
	\SetKwData{acc}{AcceptRate}\SetKwData{nIter}{nIter}
	\SetKwInOut{Input}{Input}\SetKwInOut{Output}{Output}
	
	\nonl\Fn{\ParticleMCMC{$S$, $\allObs$, $d$, $\cT$, $M$}}{
		\Input{Number of particles $S$, dataset $\allObs=\cbr{\obs vi}$, Pitman-Yor process discount parameter $\d$ and the phylogeny tree $\cT$, number of iterations $M$.}
		\Output{Posterior samples of the jump rate $\lambda$ and jump configurations $\Jump$.}
		\BlankLine
		
		Initialize $\lambda \sim \operatorname{Exp}(L^{-1})$, where $L$ is the length of the scaled tree\;
		Initialize $\Jump$, with $b_e \sim \poisson(\lambda L_e)$, $e=1,\dots,|E|$\, where $L_e$ the length of branch $e$\;
		Obtain the unbiased likelihood estimate $\widehat{p} \gets $ \pf{$\Jump, S, \allObs, d, \cT$}\;
		\For(\tcp*[f]{Gibbs update for $\lambda$; MH update for $\Jump$}){$l\gets1$ \KwTo $M$}{
			Given $\Jump$, sample $\lambda$ from \Cref{eqn:lambda_post}\;
			Given $\lambda$, propose $\Jump^*$ from $q\rgiven{\Jump^*}{\Jump}$, 
			obtain the unbiased likelihood estimate $\widehat{p^*} \gets $ \pf{$\Jump^*, S, \allObs, d, \cT$}, and
			set $A(\Jump, \Jump^*) = \frac{\widehat{p^*} }{\widehat{p}}\cdot
			\frac{ \pgiven{\Jump^*  }{\lambda} q\rgiven{\Jump}{\Jump^*} }
			{ \pgiven{\Jump}{\lambda} q\rgiven{\Jump^*}{\Jump} }$\;
			Accept with probability $\min\{1, A(\Jump, \Jump^*)\}$, in which case $\Jump\gets\Jump^*$ and $\widehat{p}\gets\widehat{p^*}$\;
		}
		\Return{posterior samples of $\lambda$ and $\Jump$}
	}
\end{algorithm}

\subsection{Updating the jump rate}
\label{sec:jump_rate}

Given $\Jump$, and with the $\mathrm{Exponential}(\rho)$ prior, 
$\lambda$ is conditionally independent of $\allObs$:
\begin{align}
\lambda \mid \Jump, \allObs
\sim
\mathrm{Gamma}\!\left(
1 + \sum_{e \in E} b_e,\,
\rho + \sum_{e \in E}\ell'(e)
\right). \label{eqn:lambda_post}
\end{align}
Here, we have used the shape--rate parameterization. Thus $\lambda$ can be sampled directly.


\subsection{Point estimation and Bayes factors} \label{sec:estimation}
Each posterior sample of $\Jump$ induces a clustering $Z=(z_v)_{v\in V}$ of the tree nodes, where $z_v$ denotes the equivalence class of $v$ under $\sim_J$. We summarize the posterior samples by minimizing posterior expected Binder loss~\citep{lau2007bayesian}.
%
Binder's loss is the number of node pairs $(u,v)$ for which clusterings $Z$ and $\widehat Z$ disagree about co-clustering:
\[
L(Z,\widehat Z)
=
\sum_{u<v}
\Big(
\mathds 1[z_u\neq z_v,\widehat z_u=\widehat z_v]
+
\mathds 1[z_u=z_v,\widehat z_u\neq\widehat z_v]
\Big).
\]
%

Define $\cL(\widehat Z)=\sum_{u<v}
\mathds 1[\widehat z_u=\widehat z_v]
\left(
\Pr(z_u=z_v\mid\allObs)-\frac12
\right)$.
Then~\citep{lau2007bayesian},
\[
\arg\min_{\widehat Z}
\mathbb E[L(Z,\widehat Z)\mid\allObs]
=
\arg\max_{\widehat Z}\cL(\widehat Z).
\]
We call this minimizer the \emph{median clustering}.
We approximate this by first estimating the co-clustering probabilities $\Pr(z_u=z_v\mid\allObs)$ for all pairs of nodes from the first half of the post-burn-in MCMC chain. Then, we evaluate $\cL(\widehat Z)$ for every clustering $\widehat Z$ visited in the second half of the chain and return the maximizer as our point estimate, along with the corresponding jump configuration.
The latter is a more informative \emph{joint} summary of the posterior over jump configurations than marginal posterior jump probabilities. 
For example, a branch and its child may each have moderate marginal posterior jump probabilities, while exhibiting strong negative dependence: a jump may typically occur on one branch or the other, but rarely on both simultaneously.

To quantify evidence for at least one jump, let $M_0$ denote the no-jump model, with $\Jump = 0$, and let $M_1$ denote the alternative model in which at least one jump occurs, with $\Jump\neq0$. The Bayes factor in favor of $M_1$ over $M_0$ is
$
K =
\frac{p(\allObs\mid M_1)}{p(\allObs\mid M_0)}=
\frac{\Pr(M_1\mid\allObs)}{\Pr(M_0\mid\allObs)}
\Big/
\frac{\Pr(M_1)}{\Pr(M_0)}.
$
The prior probabilities $\Pr(M_0)$ and $\Pr(M_1)$ are available in closed form from the Poisson prior on $\Jump$, 
while the posterior probabilities $\Pr(M_0\mid\allObs)$ and $\Pr(M_1\mid\allObs)$ are estimated from the proportions of post-burn-in MCMC samples satisfying $\Jump\equiv0$ and $\Jump\neq0$, respectively. We interpret the Bayes factor using the scale of~\citet{jeffreys1961theory}: $\log_{10}K<1$ indicates no to weak evidence for $M_1$, $\log_{10}K\in[1,2)$ strong evidence, and $\log_{10}K\geq2$ decisive evidence.

\section{Experiments} \label{sec:experiments}
We evaluate our method on synthetic and real datasets, with the primary comparison being to the \texttt{treeBreaker} approach of~\citet{ansari2016bayesian}, using its publicly available \texttt{C++} implementation\footnote{\url{https://github.com/ansariazim/treeBreaker}}. Most experiments use binary data with one observation at each leaf, the setting supported by \texttt{treeBreaker}. Our method additionally accommodates nonbinary data, observations at internal nodes, and multiple observations per node; we demonstrate the former on a post-marital residence dataset with nonbinary residence categories.

For all experiments, we rescale the tree as described in~\Cref{sec:model}, so that jumps follow a homogeneous Poisson process. Unless otherwise specified, we place an exponential prior on the Poisson rate $\lambda$ with its parameter chosen so that the prior mean number of jumps on the rescaled tree is one. We set the discount parameter to $0.5$ in most experiments; results are robust to this choice (see supplementary material). Here, we caution against very large discount parameters, which place greater prior mass on small distributional changes and can introduce identifiability issues. For the discrete observations considered here, we use a discrete base measure $H$ assigning equal probability to each outcome. We assess robustness to misspecification of the jump rate and branch length in the supplementary material.

We run 50,000 MCMC iterations for each experiment, discarding the first half as burn-in. We report total CPU runtime and the effective sample size (ESS) of the jump rate as measures of computational performance and MCMC mixing, respectively, and report ESS per second (ESS/s) as a measure that balances sampling efficiency and computational cost.

\subsection{Comparison with \texttt{treeBreaker} on synthetic data} 
\label{sec:exp-compare}

For trees with multiple moderate-sized jumps, our model can leverage dependence between distributions to outperform \texttt{treeBreaker}. We quantify jump size using (i) the total variation (TV) distance between the distributions before and after a jump and (ii) the empirical TV distance, computed from finite samples drawn from these distributions. For distributions $P$ and $Q$, $\mathrm{TV}(P,Q)=\frac{1}{2}|P-Q|_1$. In our binary setting, if $P=\mathrm{Bern}(q_1)$ and $Q=\mathrm{Bern}(q_2)$, then $\mathrm{TV}(P,Q)=|q_1-q_2|$. If $\hat q_1$ and $\hat q_2$ denote the empirical proportions of ones in samples from $P$ and $Q$, respectively, the empirical TV is $\widehat{\mathrm{TV}}=|\hat q_1-\hat q_2|$. This quantity reflects the additional variability in estimating jump sizes from finite samples.

We fix a binary tree with 200 leaves and randomly select three nested branches containing approximately 75\%, 50\%, and 25\% of the leaves in their subtrees as the jump locations. This yields four clusters with roughly equal numbers of leaves. The observation distributions follow the nested jump structure, with all three jumps having the same direction and size. Specifically, letting $p$ denote the TV distance across each jump, the Bernoulli success probabilities from top to bottom are
$
(0.5-1.5p),\quad (0.5-0.5p),\quad (0.5+0.5p),\quad (0.5+1.5p).
$

\begin{figure}
	\includegraphics[width=\linewidth]{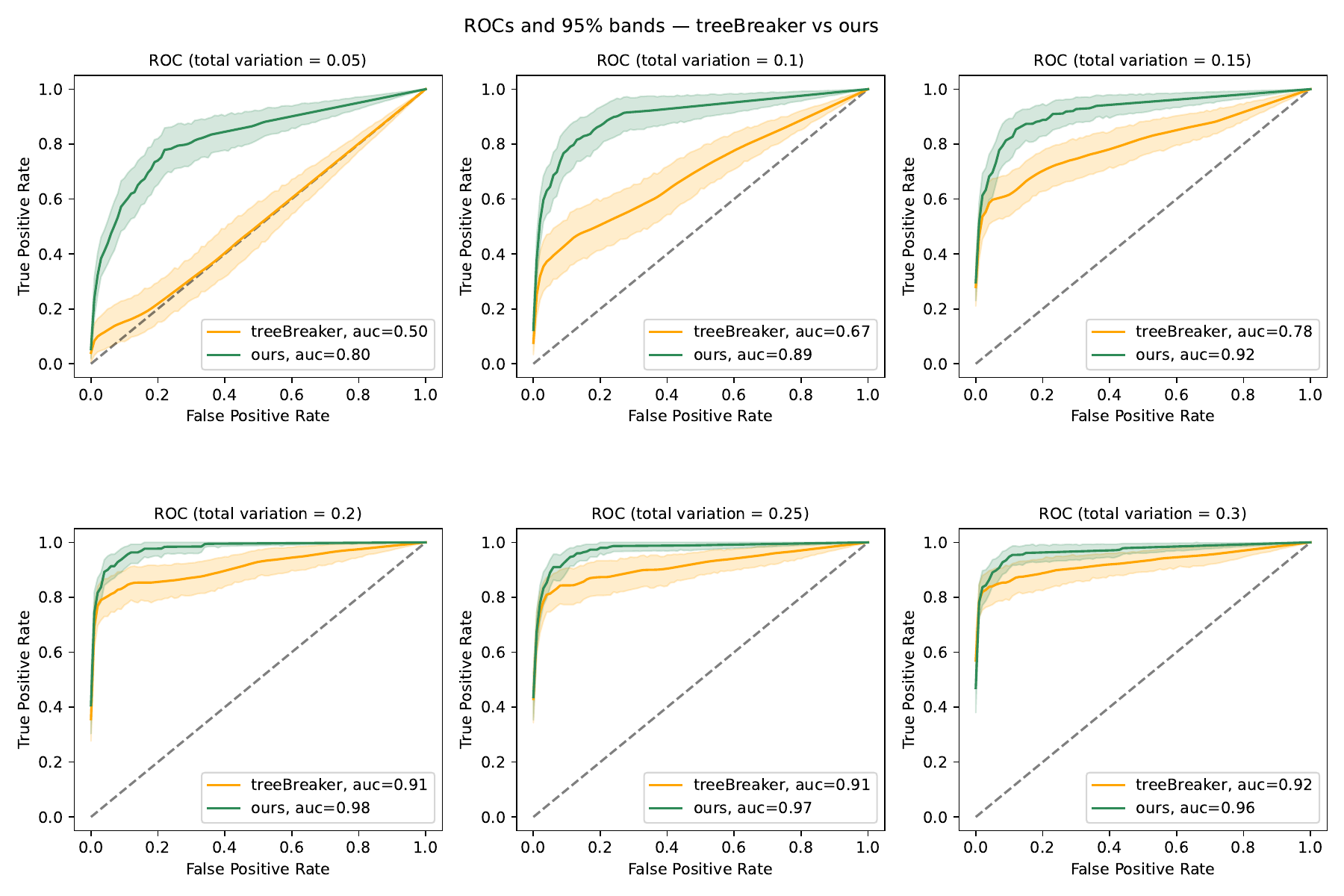}
	\caption{Average ROC curve (and shaded 95\% band) and AUC for each jump size $p$ in the synthetic comparison of \Cref{sec:exp-compare}.}
	\label{fig:syn3}
\end{figure}

We simulate 100 datasets for $p\in\{0.05,0.1,0.15,0.2,0.25,0.3\}$ and compare the true- and false-positive rates of our method and \texttt{treeBreaker} using ROC curves and AUC, averaging the ROC curves across datasets. The dashed diagonal in~\Cref{fig:syn3} represents random guessing ($\mathrm{AUC}=0.5$), with larger AUC indicating better performance. As expected, our approach performs particularly well for small jumps. For $p=0.05$, \texttt{treeBreaker} has an average AUC of 0.50, whereas our method achieves 0.8. This improvement reflects the benefit of borrowing statistical strength across dependent populations when distributions change only gradually. The advantage narrows as jumps grow easier to detect; AUCs of $0.67$ versus $0.89$ at $p=
0.10$, and $0.78$ versus $0.92$ at $p=0.15$; and both methods perform well for large jumps ($0.91$ versus $0.97$ at $p=0.25$), confirming that our method remains competitive when the signal is strong while offering substantial gains when it is weak.


We also ran~\texttt{treeSeg}~\citep{behr2020testing}: this exhibited substantially more conservative behavior, with very low per-node false-positive rates across its range of significance levels. The corresponding comparison is included and discussed in the supplementary material.

\subsection{HLA allele distribution changes in hosts in a virus phylogenetic tree} \label{sec:exp-hla}

\begin{figure}
	\includegraphics[width=\linewidth]{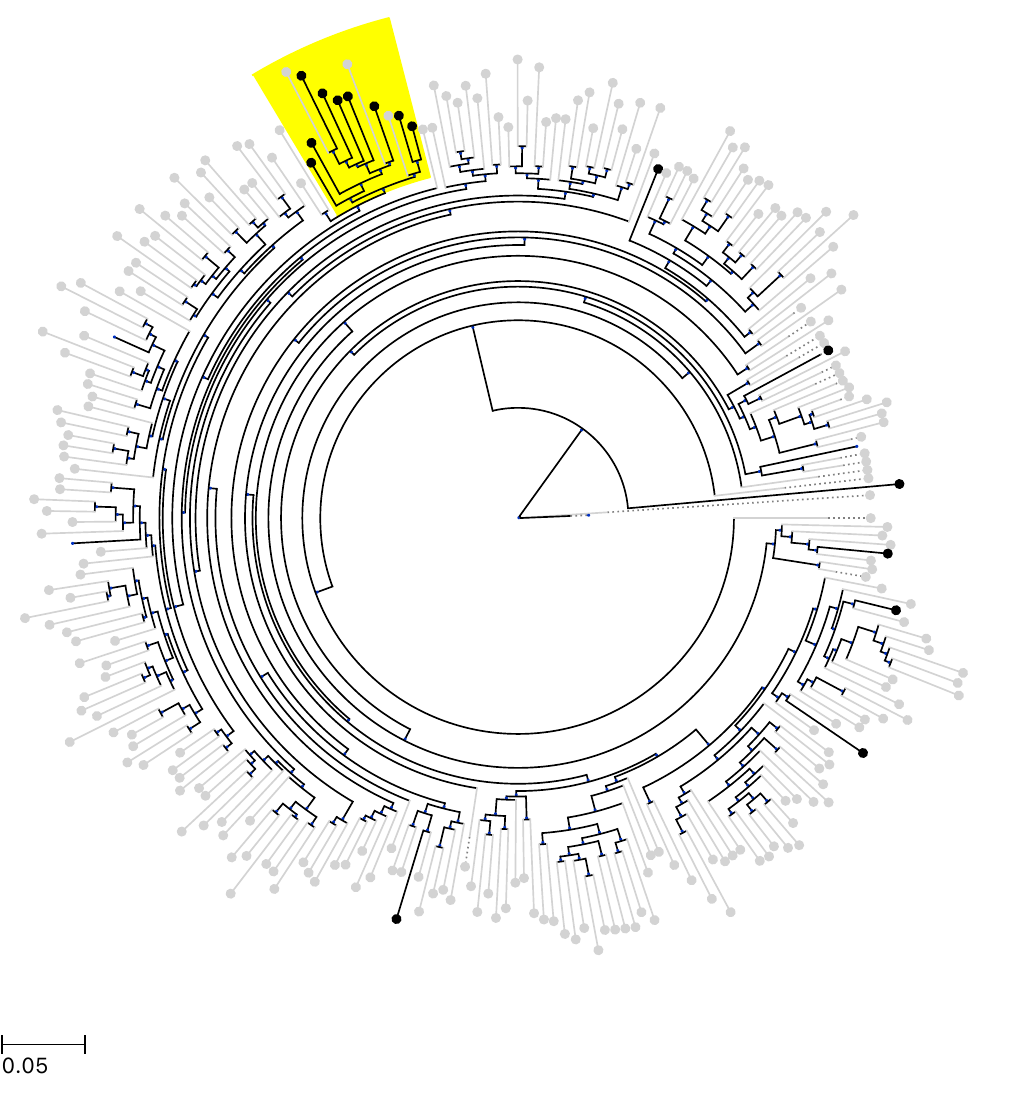}
	\caption{Results from the analysis of the HLA allele B57 dataset (\Cref{sec:exp-hla}). 
		Dots at leaf nodes represent whether the host carries allele B57 (black) or not (light grey). 
        The estimated posterior after MCMC burn-in placed no mass on the no-jump model.
		The shaded region marks the jump detected by our method, consistent with the findings of \citet{ansari2016bayesian}.
		The total runtime of our method is 2098.8s, and it produces ESS/s = 9667.}
	\label{fig:hla}
\end{figure}

We follow~\citet{ansari2016bayesian} in applying our model to the HLA allele dataset of~\citet{rousseau2008hla} (\Cref{eg:virus} in~\Cref{sec:model}). The viral whole genomes were divided into 10 segments of 1000 nucleotides, with a phylogenetic tree inferred separately for each segment, yielding 10 trees describing the evolutionary relationships among the 261 viruses. See~\citet{ansari2016bayesian} for more details. We obtained the dataset from~\url{https://www.hiv.lanl.gov/content/immunology/hlatem/study5/index.html}; this  differs slightly from that in Figure 4 of~\citet{ansari2016bayesian}, but does not affect the results.

\citet{ansari2016bayesian} applied \texttt{treeBreaker} to each HLA allele and phylogenetic tree and identified a change in the distribution of B57 on the first tree. Applying our method to the B57 observations on the first tree, we obtain a Bayes factor providing strong support for a jump. As shown in~\Cref{fig:hla}, we identify the same clade as~\citet{ansari2016bayesian}: 9 of the 12 hosts in this clade carry B57 (75\%), compared to 7 of the remaining 249 hosts (2.8\%).

\subsection{Post-marital residence distribution changes within the Uto-Aztecan language family} \label{sec:exp-pmr}

\begin{figure}
	\centering
	\includegraphics[width=.5\linewidth]{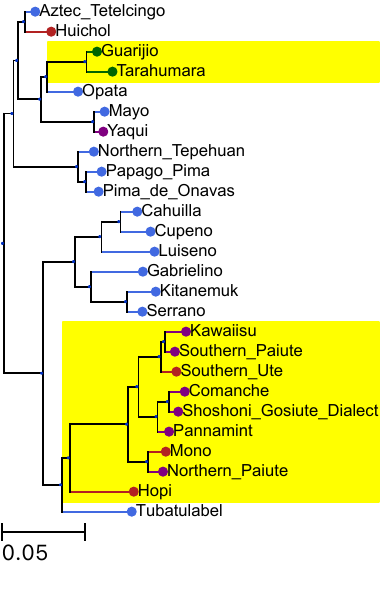}
	\caption{Results from the analysis of the post-marital residence dataset (\Cref{sec:exp-pmr}). 
		Dots at the leaf nodes indicate the four categories: patrilocality (blue), matrilocality (red), ambilocality (purple) and neolocality (green).
		The Bayes factor is 908.76, and the two detected jumps are indicated by shaded regions.  
		The runtime of our algorithm is 208.8s, and it produces ESS/s of 10.}
	\label{fig:pmr}
\end{figure}

Unlike \texttt{treeBreaker}, our method accommodates nonbinary phenotypes. We apply it to the Uto-Aztecan post-marital residence dataset of~\citet{moravec2018post} (\Cref{eg:fourcateg} in~\Cref{sec:model}), 
and obtain a Bayes factor of 908.76, providing strong evidence for distributional changes. Our method identifies two branches with jumps (\Cref{fig:pmr}).

Above both jumps (and including the root), the inferred distribution favors patrilocality, where newly married couples reside with the husband. The first jump produces a cluster at the subtree containing Guarijio and Tarahumara, in which neolocality is favored (new couples establish new residence). This pattern may reflect the poor soil in the region, which can require newly married couples to move to find productive farmland. The second jump produces a distribution strongly favoring matrilocality and ambilocality, corresponding to a transition toward desert-plains environments and associated changes in social practices.

\section{Future directions} \label{sec:discussion}

Our hierarchical nonparametric Bayesian framework opens several directions for future work. Although we focus on pure-jump evolution, the relationship between the discount parameter and the concentration of the \bnpfull\ around its centering distribution suggests a route toward modeling gradual evolution by jointly increasing the discount parameter and jump rate. We also leave extensions to continuous-valued phenotypes for future work; these could be accommodated by using nonparametric mixtures of continuous distributions, with the particle MCMC framework extending naturally to the resulting latent-cluster representation. Finally, it is of interest to extend the framework to account for uncertainty in the underlying phylogenetic tree, rather than conditioning on a single estimated tree.

\section{Acknowledgments} \label{sec:acknowledgments}
We thank Murray Cox for help interpreting results from the post-marital residence analysis. 


\bibliographystyle{abbrvnat}
\bibliography{phyloHPYP}

\label{lastpage}
\newpage
\setcounter{page}{1}

\begin{center}
	{\large\bf SUPPLEMENTARY MATERIAL}
\end{center}


As stated in the paper, our method, implemented in Python3, is available at  \url{https://github.com/varao/jumpsontree}.

	
	
	
	
	
	
\section{Additional model details}
\subsection{Description of Figure 1}
\Cref{fig:tree_and_crf}(a) clarifies the main ideas of our paper through an example tree with nine nodes and their associated distributions, the root distribution $\G_0$, five leaf distributions $\G_1,\dots,\G_5$ and three internal distributions $\G_6$, $\G_7$ and $\G_8$. 
It also shows the groups or segmentation (color blocks) introduced by the jumps (red crosses) at two branches in the tree. 
Here, the distribution at root $\G_0$ is the same as $\G_7$ and $\G_3$, so that they belong to one group. 
There are two jumps at the left branch of the root, but only one jump at the right branch of $\G_7$, which means that the left group of populations ($\G_1$, $\G_2$ and $\G_6$) has a weaker dependence on the root distribution compared to the right group ($\G_4$, $\G_5$ and $\G_8$). 

\Cref{fig:tree_and_crf}(b) demonstrates the pruned tree introduced by these jumps and the distributions associating with it, where $\G_1$ represents the distribution of the left group, $\G_0$ of the root, and $\G_4$ represents that of the right group.
Note that there is an intermediate distribution $\G'_1$ between the root and the left group $\G_1$, though  our modeling does not require us to instantiate the intermediate distributions.
In this figure, like in the rest of this work, we focus on the situation where the distributions are discrete, with the bar graphs denoting discrete distributions at observed nodes, and from which observations are drawn independently.

\Cref{fig:tree_and_crf}(c) shows a clustering of 6 observations, and   illustrates how under the Chinese restaurant franchise (Algorithm~\ref{alg:crf}), when a new table is opened in a child restaurant, it is itself seated as a customer in the parent restaurant. Here, both tables 1 and 4 (which are in different restaurants) are customers at the same table in the root (table A). 

\Cref{fig:tree_and_crf}(d) shows the resulting clustering once 4 more  observations have been generated by the CRF.

\subsection{Rescaling the tree branch lengths for a constant jump rate over depth}
We perform this through the following steps:
\begin{enumerate}
   \item partition each branch into maximal segments over which the number of branches present is constant (in the appendix, we show this reduces to sorting the nodes by depth);
    \item rescale the length of each segment by the reciprocal of that number; and then
    \item place a homogeneous rate-$\lambda$ Poisson process on the resulting rescaled tree.
\end{enumerate}

\section{Robustness checks}
\subsection{Effect of the discount parameter on the identifiability of jumps} \label{sec:exp-identifiability}
\begin{figure}[h]
	\includegraphics[width=.48\linewidth]{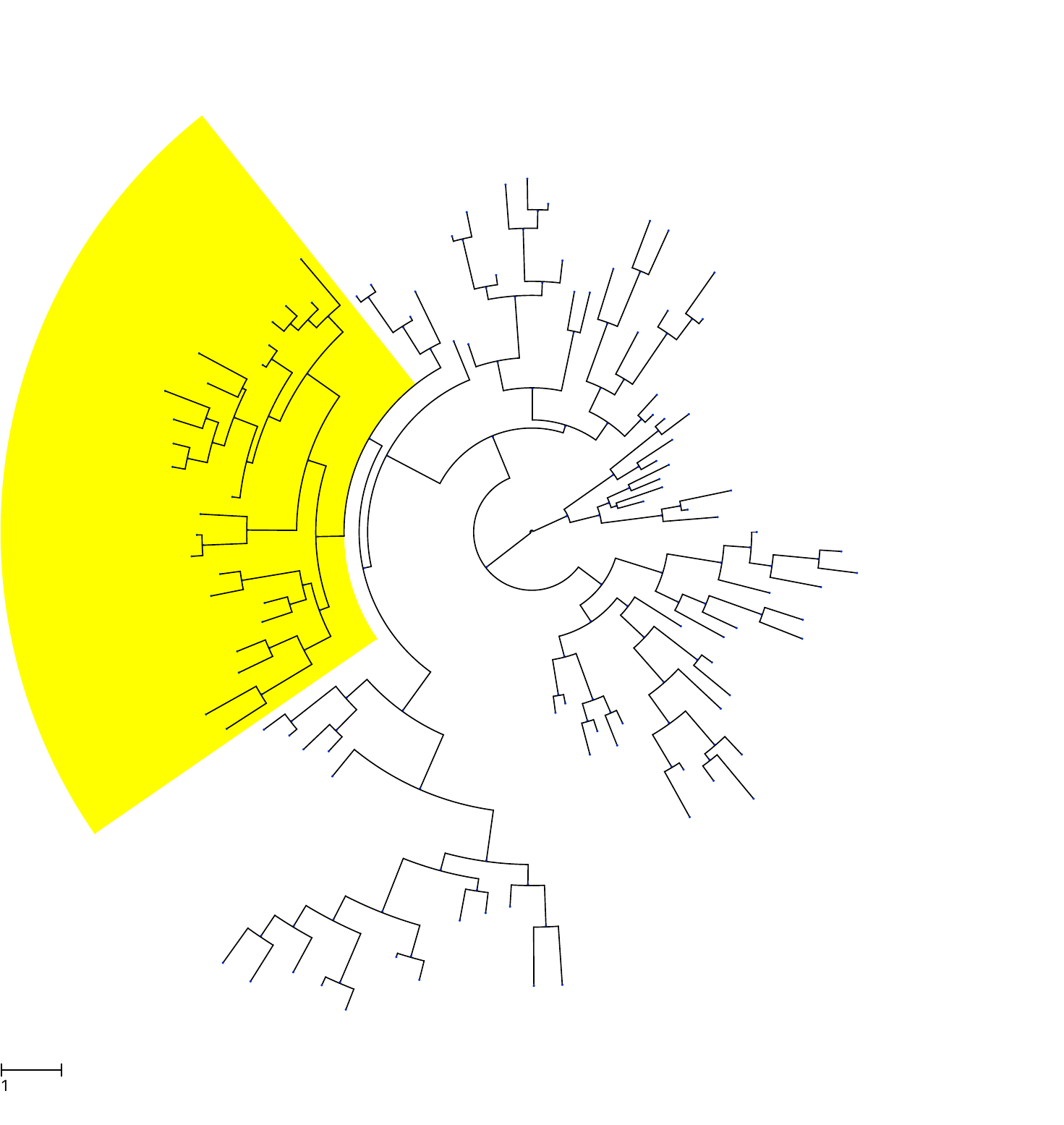}
	\includegraphics[width=.48\linewidth]{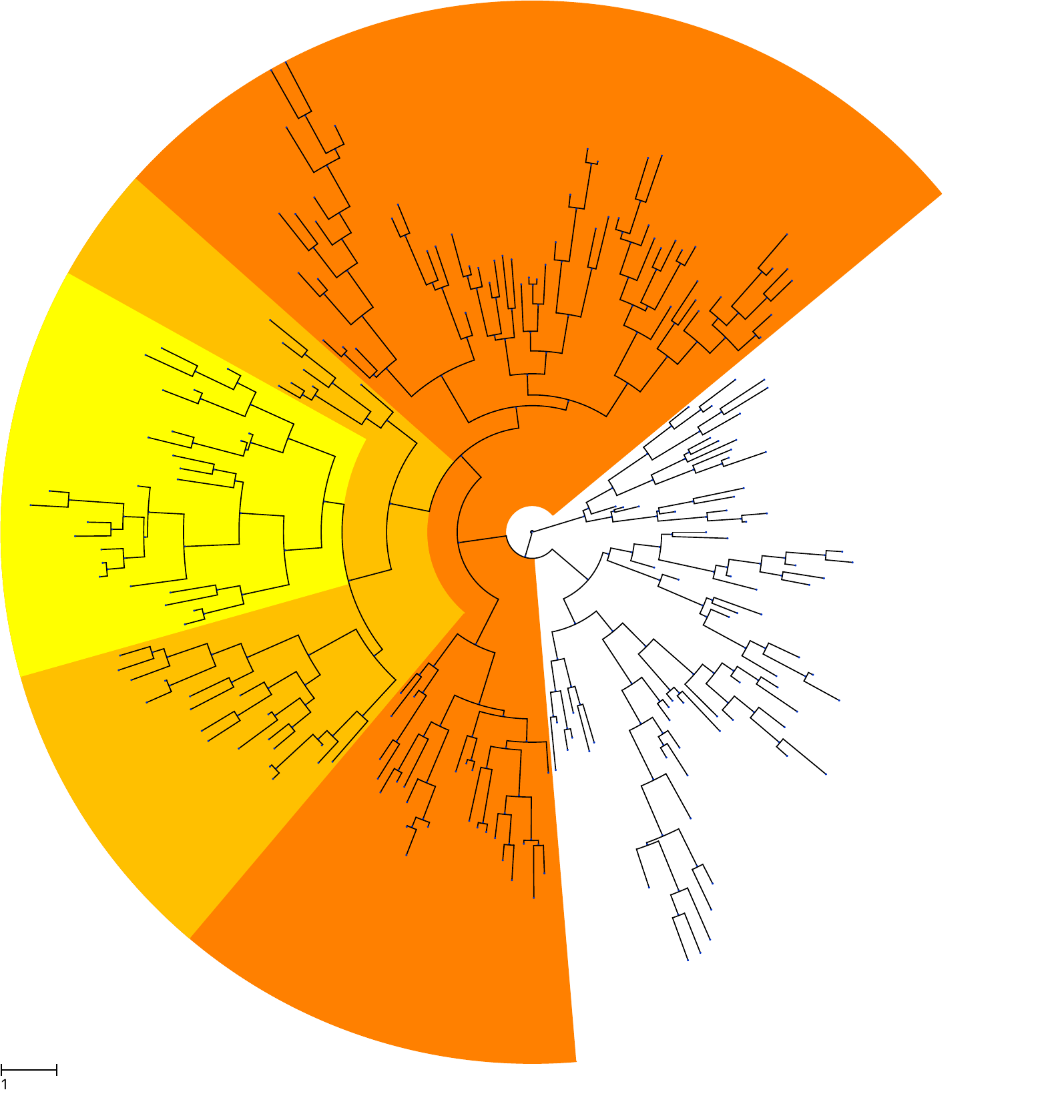}
	\caption{An example simulated tree for 
		\textbf{(left)} the first two synthetic studies in	\Cref{sec:exp-identifiability} and \ref{sec:exp-robust}, and
		\textbf{(right)} the third synthetic study in \Cref{sec:exp-compare}. 
		Color shading highlights the subtrees affected by branches with jumps. }
	\label{fig:syn-data}
\end{figure}
This experiment uses synthetic binary data to assess our model's sensitivity to hyperparameter settings.
For each replication of this experiment, we generated a random binary tree with 100 leaf nodes using the function \texttt{rtree} in the \texttt{R} package \texttt{ape} \citep{r-ape}.
A jump is uniformly assigned to a branch among those covering 10\% to 50\% of the total number of leaves in the tree, thereby ensuring the jump affects sufficient observations. 
\Cref{fig:syn-data} (left) shows an example of a simulated tree, as well as the branch selected to have a jump.
We then construct two Bernoulli distributions -- one {\em before} the jump, and one {\em after}. 
The latter is used to generate observations on leaves below the jump, and the former is used for the remaining leaves.
We set the two distributions to be symmetric about $0.5$, one having success probability $(0.5+p)$, and the other $(0.5-p)$, with $p$ the jump size, equaling the total variation between the distributions. 
We let $p$ take values  $0.1, 0.2, 0.3$ and $0.4$, and also varied the discount parameter to values in $\cbr{0.1, 0.3, 0.5, 0.7, 0.9}$. 
For each setting of the total variation and discount parameter, we ran 20 replications. 

We evaluate performance using two quantities. 
The first, which we refer to as ``target jump identified'',  takes value one when the posterior median estimate of jumps (described in \Cref{sec:estimation}) exactly matches the ground truth, that is, when the target branch and only the target branch is present in the posterior median.  
The second uses the MCMC samples to compute the Bayes factor of the model with at least one jump to the null model with no jump (see \Cref{sec:estimation} for details).
The Bayes factor shows how confident we are of the existence of jumps, with  a Bayes factor greater than $10^2$ suggesting decisive evidence towards the existence of jumps in the tree.


\begin{figure}[h] 
 	\includegraphics[width=.48\linewidth]{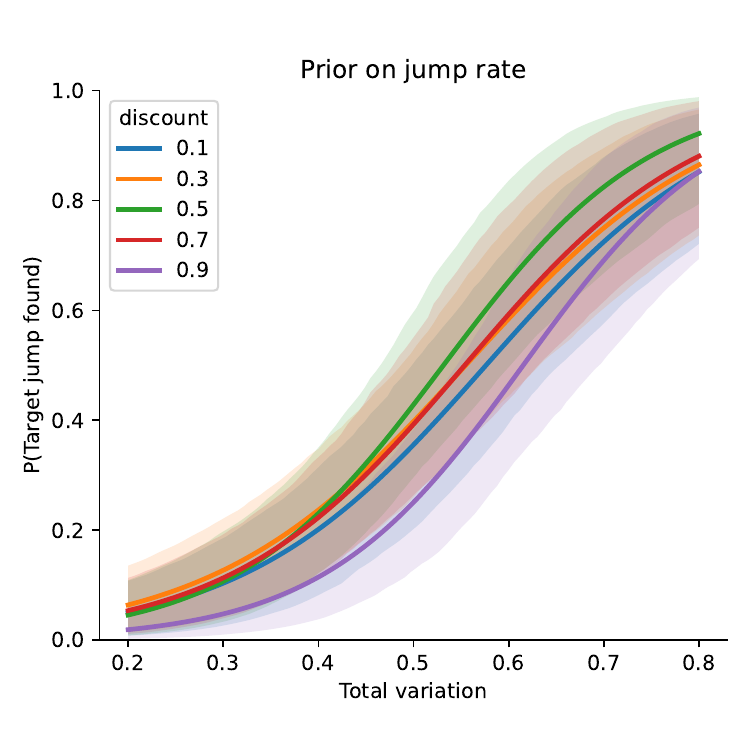}
 	\includegraphics[width=.48\linewidth]{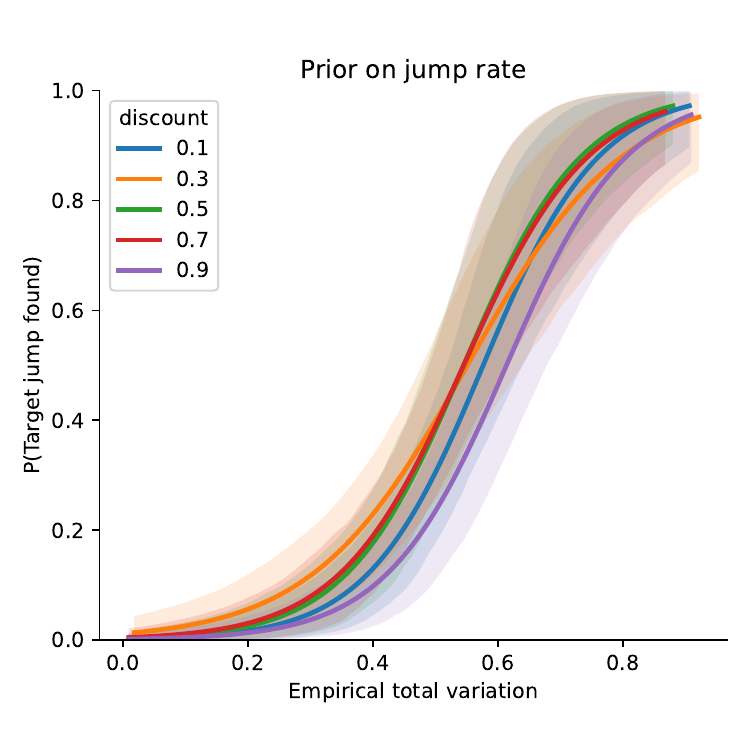}
 	\caption{ Estimated probability of identifying the target jump (and the 95\% percentile bands) versus the total variation. The probability is estimated by fitting a logistic regression to the indicator of identifying the target jump. 
		\textbf{(right)} The same plot, now against the empirical total variation.} 
	\label{fig:syn1-jr}
\end{figure}
\begin{figure}[h]
\begin{minipage}{.6\linewidth}
 	\includegraphics[width=.98\linewidth]{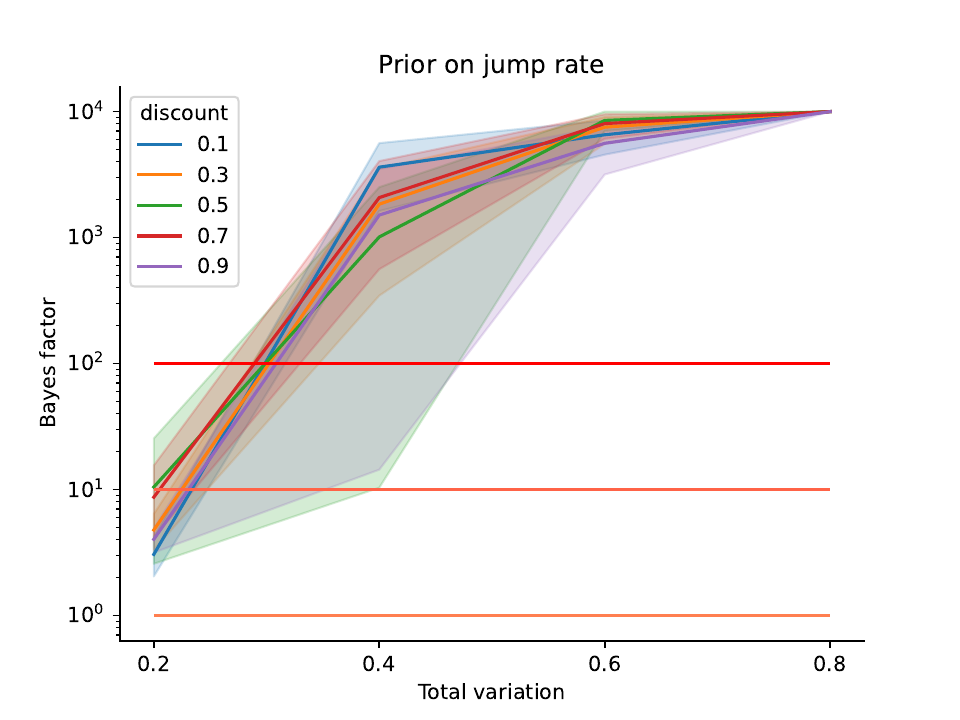}
  	
\end{minipage}
  \begin{minipage}{.38\linewidth}
	\caption{ Bayes factors (mean and the 95\% percentile bands) versus the total variation. The Bayes factor is truncated at $10^4$ to make the plot. The horizontal lines mark the conventional decision boundaries of Bayes factor described in \Cref{sec:estimation}.	}
  \label{fig:syn1-jr_bf}
 \end{minipage}
\end{figure}


\Cref{fig:syn1-jr} summarizes the results, with the left panel plotting the probability that the target jump is identified against the  total variation distance. Here, we smoothen the results by fitting a logistic regression model on the indicator variable ``target jump identified'', with the total variation as the covariate variable. When the total variation is small (0.1 and 0.2), there is little evidence in the data to support the existence of a jump, and as we expect, the probability of identifying the jump is small. 
For large jump sizes, this probability increases significantly, though doesn't reach one for all settings of the discount parameter. The right panel plots performance as a function of the empirical TV, rather than TV, allowing us to control for fluctuations about the population TV because of finite sample effects. With this, we see that probabilities approach 1 for large jumps for all settings of the discount parameter. 
Different choices of the discount parameter do not strongly impact the ability to determine the existence of jumps, although a very large discount parameter can harm the ability to correctly locating the jump in the tree. 
This is expected, since with a high discount parameter, the model favors many jumps with smaller sizes, resulting in false positives. 
\Cref{fig:syn1-jr_bf} plots Bayes factors 
against TV, and again, we see this increases with jump size. The Bayes factor is also relatively insensitive to the discount parameters. 
In practice, we recommend discount parameter values that are not too small or too large, and going forward, fix $d= 0.5$.

\begin{figure}[h]
 	\includegraphics[width=.45\linewidth]{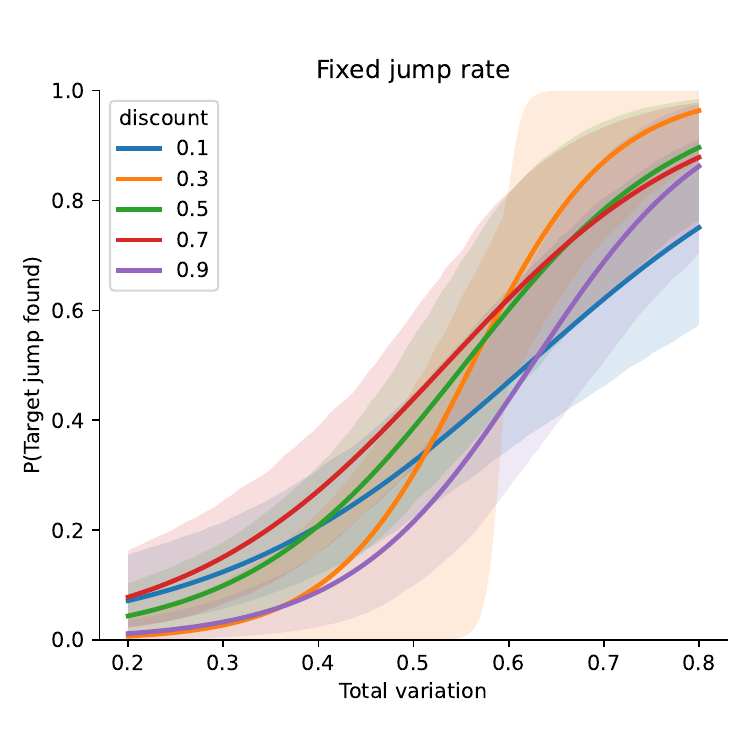}
 	\includegraphics[width=.45\linewidth]{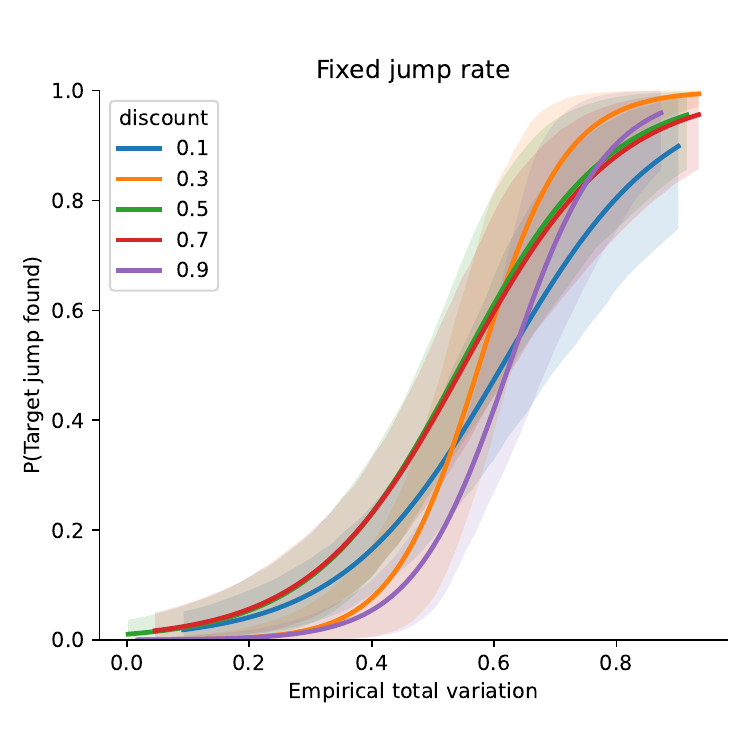}

\begin{minipage}{.6\linewidth}
 	\includegraphics[width=.98\linewidth]{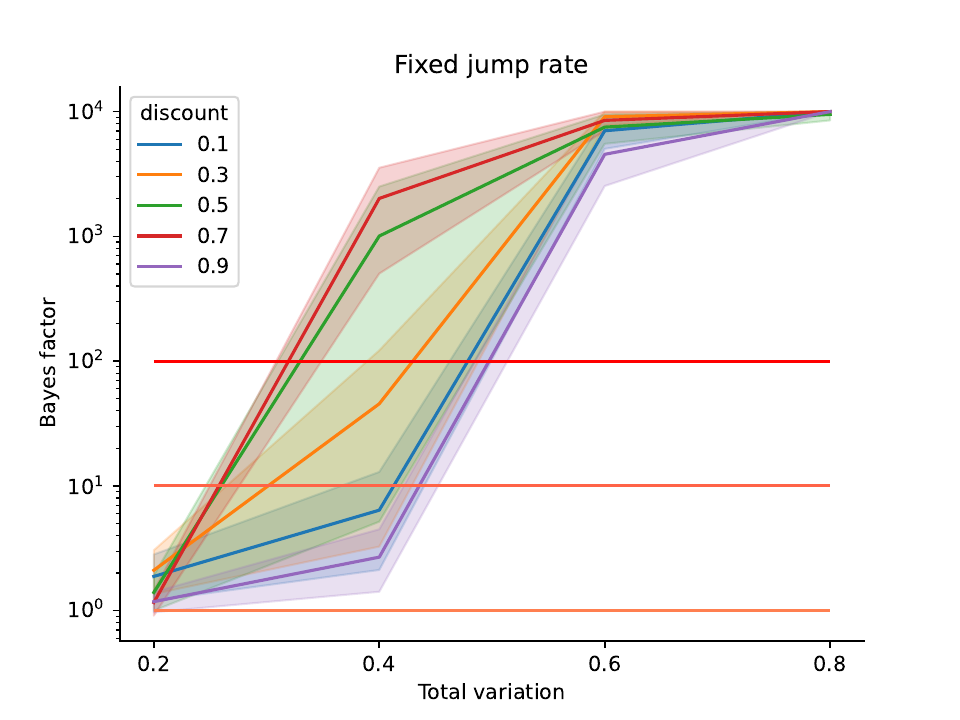}  	
\end{minipage}
  \begin{minipage}{.38\linewidth}
  \caption{Synthetic study results of \Cref{sec:exp-identifiability} when the jump rate is fixed. The top row plots the probability of detecting the target jump, and the bottom row plots Bayes factors. The detailed figure descriptions are as in \Cref{fig:syn1-jr} and   \Cref{fig:syn1-jr_bf}.}
 \label{fig:syn_fixjr}
 \end{minipage}
\end{figure}
We further evaluate our model when the jump rate is fixed to have  one jump in the tree on average. 
\Cref{fig:syn_fixjr} shows the results. 
Similar to when the jump rate is learnt, our algorithm performs well when there is a large change in the distribution before and after the jump.

\subsection{Robustness to misspecification of jump rate} \label{sec:exp-robust}
\begin{figure}
 	\includegraphics[width=.48\linewidth]{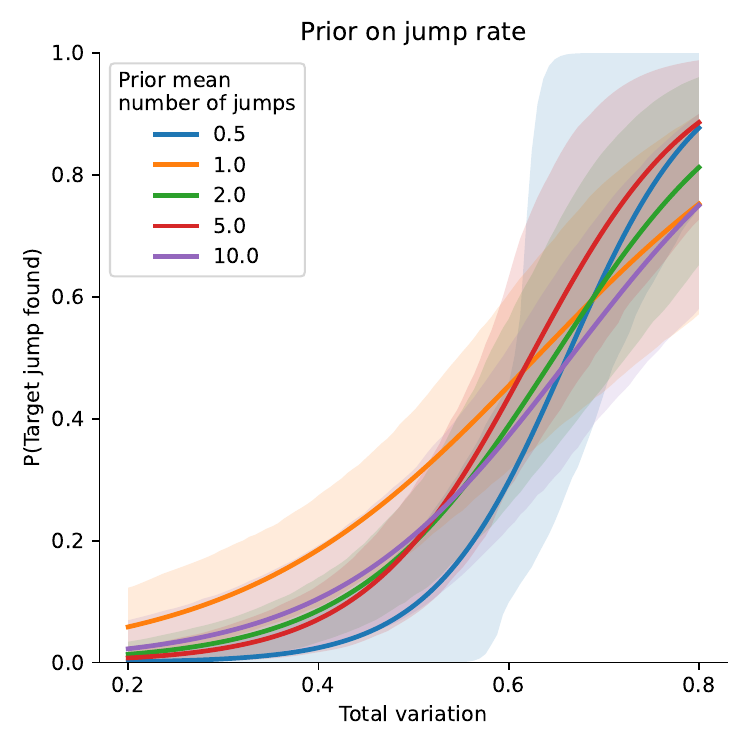}
 	\includegraphics[width=.48\linewidth]{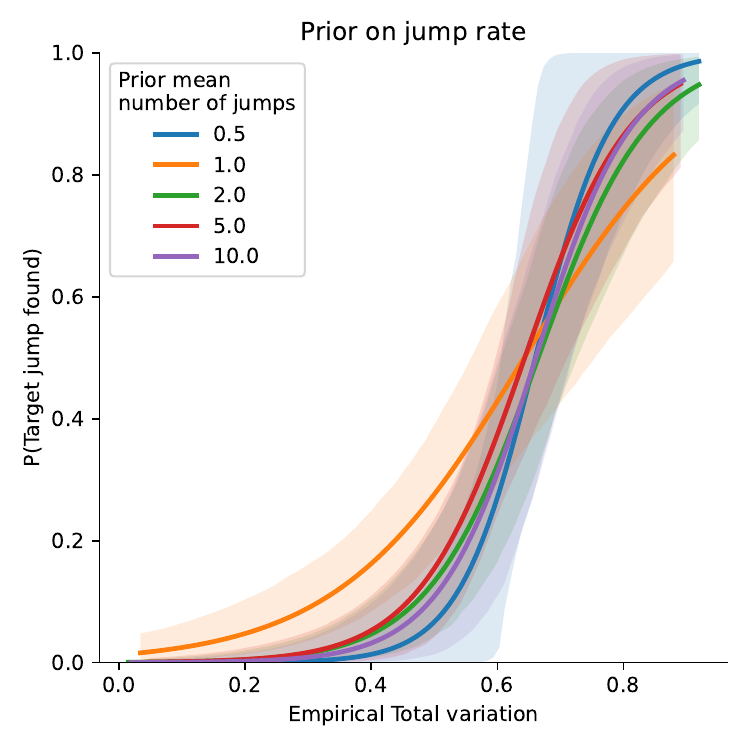}
	\caption{Synthetic study results of \Cref{sec:exp-robust}, showing the effect of different settings of the prior on the jump rate on the probability of detecting the target jump. The detailed figure description is as in \Cref{fig:syn1-jr}.}
	\label{fig:syn2-jr}
\end{figure}

Next, we study how sensitive our approach is to misspecification of the jump rate. 
Our setup is similar to the previous \Cref{sec:exp-identifiability}, except 
here, we fix the discount parameter to be 0.5.
We place an exponential prior on the Poisson jump rate, and investigate performance as we vary its mean between 0.5, 1, 2, 5, or 10 jumps in the tree. {In all cases, the ground truth had exactly one jump in the tree.}
The results are shown in \Cref{fig:syn2-jr}, and it is clear that the choice of prior on the rate does not have a significant impact on the probability of identifying the target branch.
Thus, when learning the jump rate, our model is very robust to potential misspecification of the prior on jump rate. 


\begin{figure}
 	\includegraphics[width=.48\linewidth]{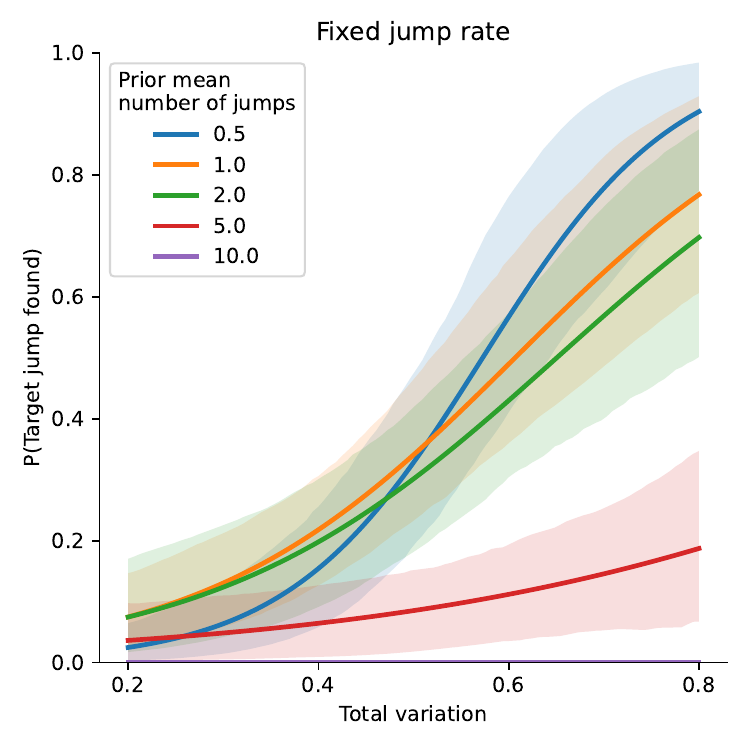}
 	\includegraphics[width=.48\linewidth]{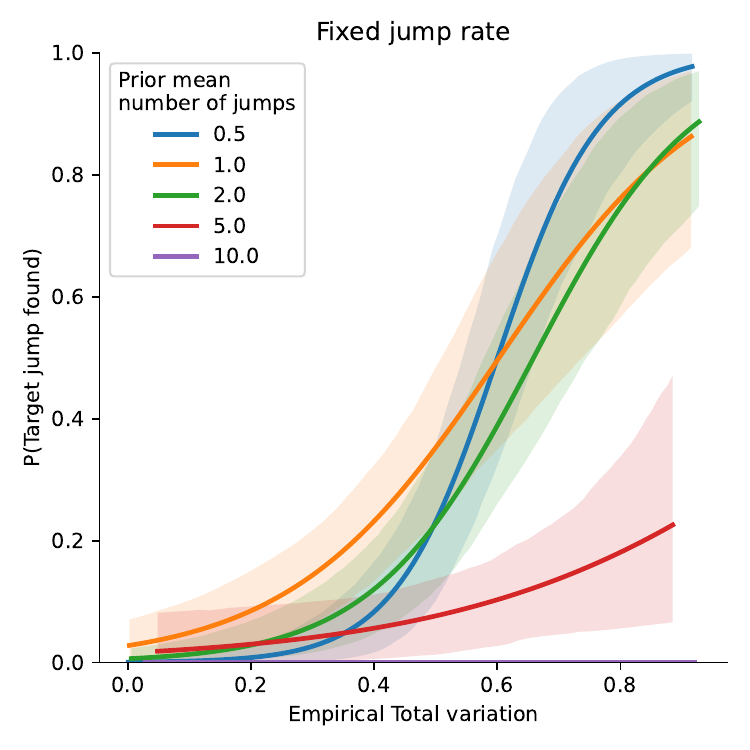}
	\caption{Synthetic study results of \Cref{sec:exp-robust}, showing the effect of different settings of the jump rate on the probability of detecting the target jump. The detailed figure description is as in \Cref{fig:syn1-jr}.}
	\label{fig:syn2-jr-fix}
\end{figure}
\begin{figure}
	\includegraphics[width=.48\linewidth]{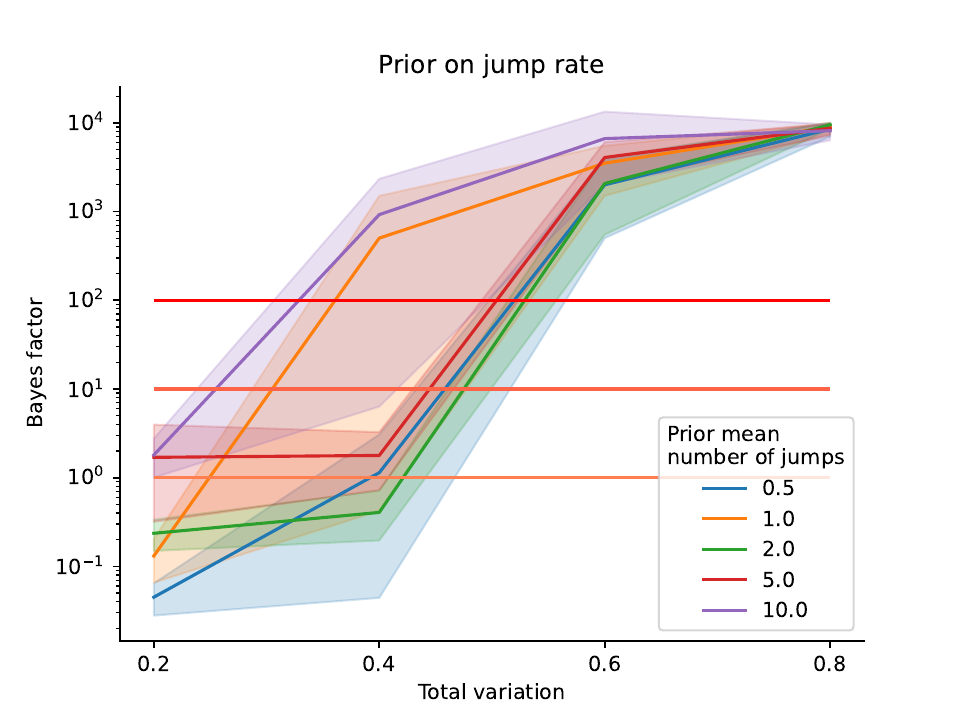}
 	\includegraphics[width=.48\linewidth]{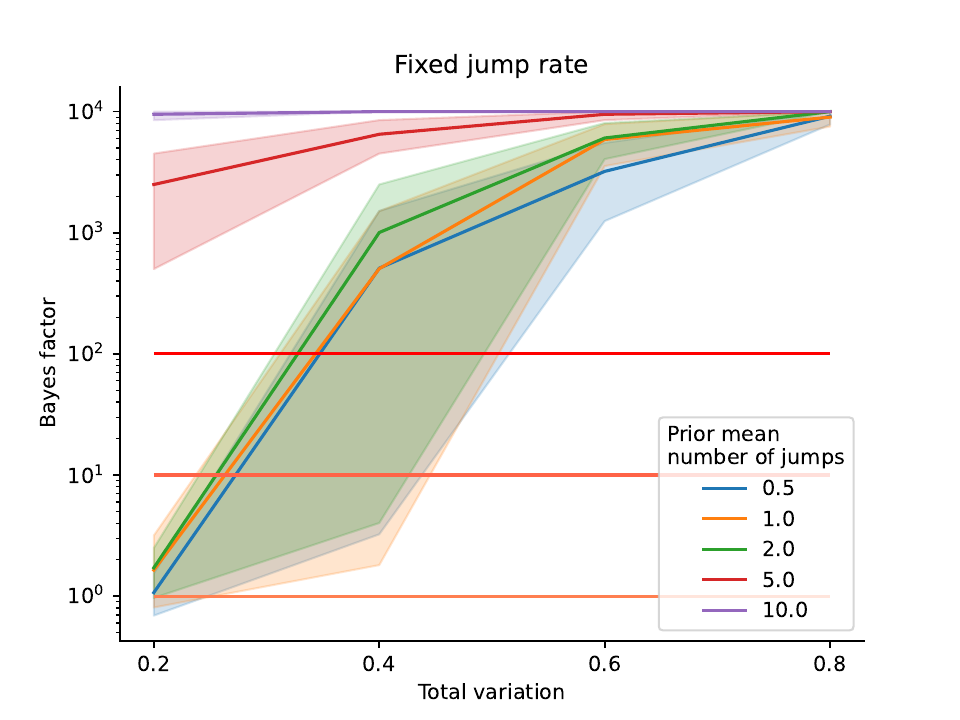}
	\caption{Synthetic study results of \Cref{sec:exp-robust}, showing the effect of different settings of the prior on the jump rate itself (left) and jump rate (right) on Bayes factors. The detailed figure description is as in \Cref{fig:syn1-jr_bf}.}
	\label{fig:syn2-bf}
\end{figure}
We also investigate performance when the jump rate is fixed to have, on average 0.5, 1, 2, 5, or 10 jumps in the tree. \Cref{fig:syn2-jr-fix} shows the results: 
 here, fixing the jump rate incorrectly has a stronger impact on performance, and with a high prior number of jumps (5 or 10), the probability of correctly detecting only the target jump drops steeply (the curve when the rate is set to 10 is indistinguishable from 0). 
We see a similar story looking at Bayes factors in the right panel {of~\Cref{fig:syn2-bf}},  with a high prior number of jumps (5 or 10), the Bayes factor stays very high even when the data provides weak support to the existence of jumps. 
Taking this into account, we recommend uninformative Bayesian inference on the jump rate {when there is no strong prior knowledge of the number of jumps in the tree.}

\subsection{Robustness to branch lengths}
In this experiment, we reran our method on the synthetic datasets of the main paper at $p=0.15$, with all branch lengths set to a common constant, removing branch-length information. Specifically, the original branch lengths used to generate the data took values in the interval $(0,1]$, and we set them all to $0.1$ (a natural intermediate value), before applying the depth-based rescaling of~\Cref{sec:poiss_tree}. The average AUC dropped from 0.92 to 0.80 across the replicates. Notably, the $0.80$ attained without branch-length information is comparable to {\tt treeBreaker}'s AUC of 0.78 with full branch lengths. We conclude that, while our method is sensitive to branch lengths by design — they determine the prior probability of a jump on each branch — it remains useful in settings where branch lengths are unavailable.
\begin{figure}
	\includegraphics[width=\linewidth]{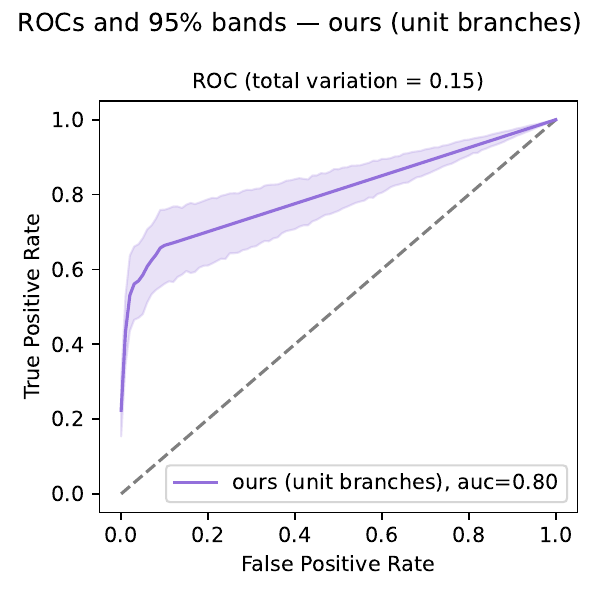}
	\caption{ROC plot for our method with tree branches set to 0.1}
	\label{fig:roc_unit}
\end{figure}

 \subsection{Comparison with {\tt treeSeg}}
 \begin{figure}
	\includegraphics[width=.98\linewidth]{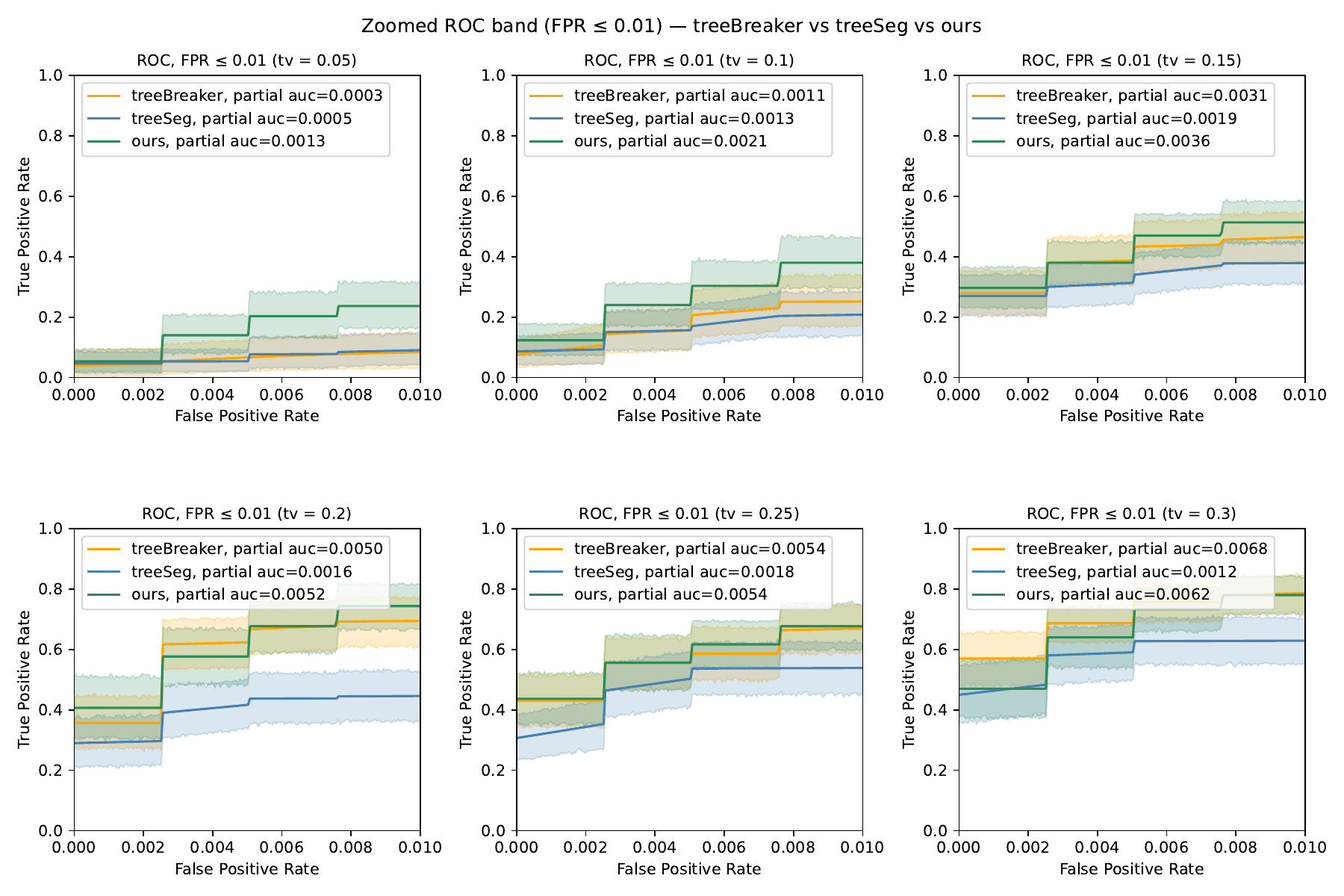}
	\caption{ROC plots of our method, treeBreaker and treeSeg, zoomed into the operating region of treeSeg}
	\label{fig:treeseg}
\end{figure}
In~\Cref{fig:treeseg}, we include a comparison with {\tt treeSeg} from~\citet{behr2020testing}. {\tt treeSeg} uses a significance parameter $\alpha$ that provides a tree-wide false-positive guarantee, controlling the probability of incorrectly detecting a spurious segment anywhere on the tree. To construct a ROC curve, we ran treeSeg across a grid of $\alpha$ values ranging from $0.001$ to $0.95$.

{\tt treeSeg} exhibited a very low per-node false-positive rate (0.3\% even at $\alpha = 0.9$), consistent with the conservative behavior reported in the original study. The latter reported a 1.4\% false-positive rate at $\alpha = 0.1$ as a per-tree measure, that is, the proportion of trees with no true signal in which at least one edge was incorrectly flagged. Our 0.3\% is a per-edge measure, calculated as the fraction of true non-jump edges incorrectly detected across all replicates.

We note that for large values of $\alpha$, {\tt treeSeg} did not return within the allotted runtime for some datasets. We terminated such runs after three minutes and excluded them from the corresponding ROC estimates.


\subsection{MCMC mixing diagnostics}

\begin{figure}
	\includegraphics[width=.48\linewidth]{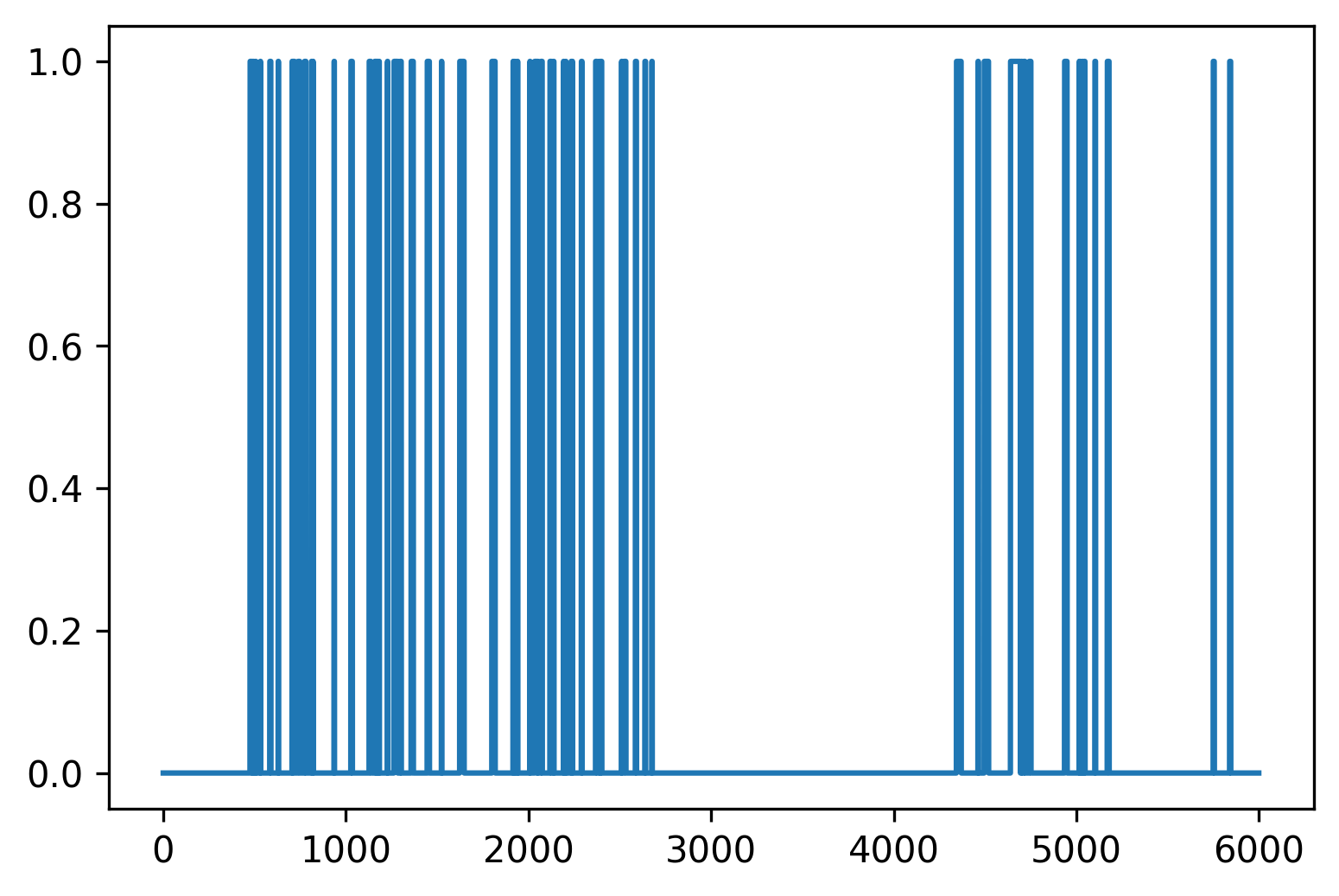}
	\caption{Trace-plot showing the number of jumps on the 'true' branch  for a dataset with \(p=0.3\) from the synthetic experiments in the main paper. ESS: 5.14 per 100 samples}
	\label{fig:mcmc_trace}
\end{figure}

\begin{figure}
	\includegraphics[width=\linewidth]{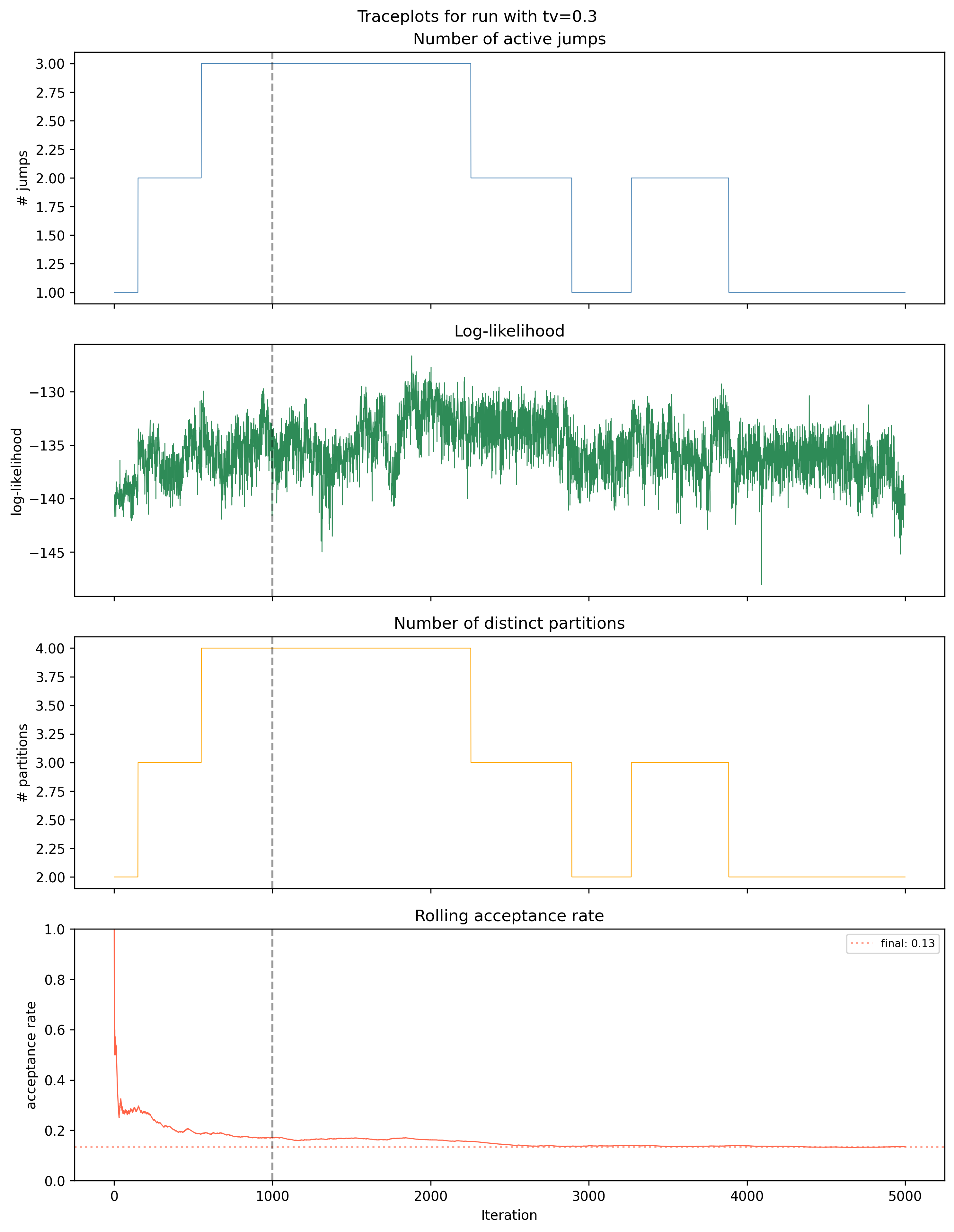}
	\caption{Trace plots of the number of jumps, log-likelihood, number of tree segmentations, and running acceptance rate for a dataset with \(p=0.3\) from the synthetic experiments in the main paper.}
	\label{fig:mcmc_trace2}
\end{figure}

\pagebreak

\setcounter{page}{24}

\end{document}